\documentclass[fleqn,10pt]{wlscirep}
\pdfoutput=1

\usepackage{amsmath}
\usepackage{amssymb}
\usepackage{mathrsfs}
\usepackage{bm}
\usepackage{url}
\usepackage{csquotes}
\MakeOuterQuote{"}
\usepackage{authblk}
\usepackage{lmodern}
\usepackage{chapterbib}

\def\vec#1{\bm{#1}} 
\newcommand{\mrm}[1]{\mathrm{#1}}
\newcommand{\tr}{\operatorname{tr}}
\newcommand{\Tr}{\operatorname{Tr}}

\newcommand{\rmd}{\mathrm{d}}

\newcommand{\be}{\begin{equation}}
\newcommand{\ee}{\end{equation}}
\newcommand{\ba}{\begin{align}}
\newcommand{\ea}{\end{align}}

\def\<{\langle}  
\def\>{\rangle}  

\newcommand{\dket}[1]{| #1\>\!\>}

\newcommand{\dbra}[1]{\<\!\< #1|}

\newcommand{\inner}[2]{\<#1|#2\>}
\def\outer#1#2{|#1\>\<#2|}       
\newcommand{\dinner}[2]{\<\!\< #1| #2\>\!\>}

\newcommand{\douter}[2]{| #1\>\!\>\<\!\< #2|}

\newcommand{\norm}[1]{\parallel\!#1\!\parallel}

\newcommand{\barcal}[1]{\bar{\mathcal{#1}}}

\newcommand{\bid}{\bar{\mathbf{I}}}
\newcommand{\ob}[1]{\mathbf{#1}} 

\def\eqref#1{\textup{(\ref{#1})}}  
\newcommand{\eref}[1]{Eq.~\textup{(\ref{#1})}}
\newcommand{\Eref}[1]{Equation~\textup{(\ref{#1})}}
\newcommand{\esref}[1]{Eqs.~\textup{(\ref{#1})}}

\newcommand{\fref}[1]{Fig.~\ref{#1}}
\newcommand{\Fref}[1]{Figure~\ref{#1}}

\newcommand{\cref}[1]{Conjecture~\ref{#1}}
\newcommand{\Cref}[1]{Conjecture~\ref{#1}}

\newcommand{\rcite}[1]{Ref.~\cite{#1}}
\newcommand{\rscite}[1]{Refs.~\cite{#1}}

\title{Information complementarity: A new paradigm for decoding quantum incompatibility}
\author[1,*]{Huangjun Zhu}
\affil[1]{Perimeter Institute for Theoretical Physics, Waterloo, On N2L 2Y5,
Canada}
\affil[*]{hzhu@pitp.ca}

\begin{abstract}

The existence of observables that are incompatible or not jointly measurable is a characteristic feature of quantum mechanics, which lies at the root of a number of nonclassical phenomena, such as uncertainty relations, wave--particle dual behavior, Bell-inequality violation, and contextuality.
However,  no intuitive  criterion  is available for determining  the compatibility of even two (generalized) observables, despite the overarching importance of this problem  and intensive efforts of many researchers.
Here we introduce an information theoretic  paradigm together with an intuitive geometric picture for decoding incompatible observables, starting from two simple ideas:   Every  observable can only provide limited  information and   information is monotonic  under data processing. By virtue of quantum estimation theory, we introduce a family of universal criteria for detecting incompatible observables and a natural measure of incompatibility, which are applicable to arbitrary number of arbitrary observables. Based on this framework, we derive a family of universal  measurement uncertainty relations,
provide a simple information theoretic explanation of quantitative wave--particle duality, and offer new perspectives for understanding Bell nonlocality, contextuality, and quantum precision limit.

\end{abstract}
\begin{document}

\flushbottom
\date{\today}
\maketitle

\thispagestyle{empty}

\begin{cbunit}
\section*{Introduction}
Observables that are incompatible or not jointly measurable play a fundamental role in quantum mechanics and quantum information science.
Profound consequences of  incompatible observables were realized soon after the inception of quantum theory by Heisenberg in a seminal paper~\cite{Heis27}, from which originated the idea of uncertainty relations~\cite{BuscHL07,WehnW10}.  Around the same time, Bohr conceived the idea of the complementarity
principle \cite{Bohr28}.  A vivid manifestation  is the famous  example of wave--particle duality \cite{Bohr28, WootZ79,SculEW91,JaegSV95, Engl96,BuscS06}. In addition, incompatible observables are intimately connected to  Bell nonlocality \cite{Bell64,ClauHSH69,BrunCPS13},   Einstein--Podolsky--Rosen (EPR) steering \cite{WiseJD07,UolaMG14,QuinVB14}, contextuality \cite{KochS67,LianSW11,AbraB11,CabeSW14}, superdense coding~\cite{Cole13}, etc. The implications of incompatibility  have never been fully explored, as reflected in  a recent heated debate on as well as resurgence of interest in measurement uncertainty and error-disturbance relations \cite{Ozaw03,BuscHL07,BuscLW13,BuscHOW14}.

Most existing  literature on incompatible observables focus on two sharp observables (those represented by self-adjoint operators), partly due to the lack of a suitable tool for dealing with more  observables or generalized observables (those described by probability operator measurements, also known as positive operator valued measures). With the advance of  quantum information science and technologies, it is becoming increasingly important to consider more general situations. Detection  and characterization of  incompatible observables is thus of paramount importance.  There exist a number of different notions characterizing the compatibility relations among quantum observables; prominent examples include commutativity, nondisturbance, joint measurability, and coexistence \cite{HeinW10,ReebRW13}.
For sharp observables, all four notions are equivalent~\cite{Neum55}. For generalized observables, however, all of them are inequivalent:  observables satisfying a former relation also satisfy a latter relation but not vice versa in general \cite{HeinW10,ReebRW13}.

 Among the four notions of compatibility mentioned above, joint measurability is distinguished by its close relation  to Bell nonlocality \cite{Bell64,ClauHSH69,BrunCPS13} and EPR steering \cite{WiseJD07,UolaMG14,QuinVB14}. In particular,  a set of observables is not joint measurable if and only if it can be used to reveal EPR steering \cite{UolaMG14,QuinVB14}.  In the rest of this paper, we shall focus on the compatibility relation  captured by the notion of  joint measurability.
Although the compatibility of a set of observables can be determined by semidefinite programming \cite{WolfPF09}, the computational complexity increases exponentially with the number of observables. In addition, existing algorithms provide little intuition as to why a set of observables is compatible or not. Actually,  no intuitive criteria is known for determining the compatibility of even two generalized observables, except for a few special cases, such as two binary observables in the case of a qubit \cite{Busc86,BuscS06,StanRH08,BuscS10, YuLLO10}. What is worse, most known criteria are derived with either  brute force or ad hoc  mathematical tricks, which offer little insight even if the conclusions are found. In this work we aim to change this situation.

In addition to the detection of incompatibility, quantification of  incompatibility is also of paramount importance. Incompatibility measures are closely related to quantitative wave--particle duality relations \cite{WootZ79,SculEW91,JaegSV95, Engl96,BuscS06} and measurement  uncertainty relations \cite{Muyn00, BuscHL07}. In this context, it is instructive to distinguish two different uncertainty relations concerning  state preparations and measurements, respectively, as clarified  in \rcite{Muyn00}.
The traditional uncertainty relation, encoded in the Robertson inequality \cite{Robe29}, characterizes preparation uncertainty. Although this is well known as the Heisenberg uncertainty relation, it is different from the measurement uncertainty relation Heisenberg had in mind~\cite{Heis27,Muyn00}. Also, most other uncertainty relations  known in the literature belong to this  type, including many entropic uncertainty relations \cite{WehnW10}. By contrast, few works have studied
measurement uncertainty relations for a long time; notable exceptions include  \rscite{MartM90,MartM90T}.
 Recently, increasing attention has been directed to  measurement uncertainty relations and incompatibility measures \cite{Ozaw03,BuscHL07,BuscLW13,BuscHOW14, BuscHSS13}. However, most works are tailored to deal with restricted scenarios, such as von Neumann observables or two generalized observables. More powerful tools are needed to deal with general settings.

In this work we propose a new paradigm for detecting and characterizing  incompatible observables. Our framework is based on simple information theoretical ideas and quantum estimation theory \cite{Hels76book,Hole82book}. The \emph{Fisher information} underpinning our study turns out to be more effective than Shannon information in capturing the compatibility relations among different observables.
In particular,  we introduce a family of universal criteria for detecting incompatible observables and a natural measure of incompatibility, which are applicable to arbitrary number of arbitrary observables.  Based on  this
framework, we derive a family of universal  measurement uncertainty relations, which substantially improve over known uncertainty relations in
terms of the scope of applicability. We also
provide a simple information theoretic explanation of quantitative wave--particle duality and derive complementary relations for more than two complementary observables.  In addition, our work
offers new perspectives for understanding Bell nonlocality, EPR steering, contextuality, and quantum precision limit.

\section*{Results}
\subsection*{Simple ideas}
Our approach for detecting and characterizing  incompatible observables is based on two simple information theoretic ideas:  (1) every observable or measurement  can only provide limited  information  and (2)  information is monotonic under data processing.  The joint observable of a set of   observables  is at least as informative as each marginal observable with respect to any reasonable information measure. A set of  observables  cannot be compatible if  any hypothetical joint measurement would provide too much information. These ideas are  general enough for dealing  with arbitrary number of arbitrary observables. Furthermore, they are  applicable not only to the  quantum theory, but also to generalized probability theories \cite{Hard01, Barr07}. For concreteness, however, we shall focus on the quantum theory.

Although  information measures are not a priori unique, we find the Fisher information~\cite{Fish25}  is a perfect choice  for our purpose. Compared with Shannon information commonly employed in relevant studies, Fisher information is usually quantified by a matrix instead of a scalar and is  more suitable for  characterising different types of information provided by different observables. In particular, Fisher information is more effective in capturing the information tradeoff among incompatible observables. In addition,  many tools in quantum estimation theory \cite{Hels76book,Hole82book} can be applied to derive incompatibility criteria and measures in a systematic way instead of relying on ad hoc mathematical tricks, as is the case in most existing studies. Consequently, the incompatibility criteria and measures we derive are more intuitive and have a wider applicability.

Suppose the states of interest are parametrized by a set of parameters denoted collectively by $\theta$. A  measurement is determined by a family of probability distributions $p(\xi|\theta)$
parametrized  by  $\theta$. The Fisher information matrix associated with the measurement is given by
\begin{equation}
\begin{aligned}
I_{jk}(\theta)&=\sum_\xi p(\xi|\theta)\frac{\partial \ln
p(\xi|\theta)}{\partial \theta_j}\frac{\partial \ln
p(\xi|\theta)}{\partial \theta_k}.
\end{aligned}
\end{equation}
Its significance is reflected in the famous \emph{Cram\'er--Rao
bound}: the  mean square error (MSE) matrix of any unbiased estimator of $\theta$ is bounded from below by the inverse Fisher information matrix (see supplementary information).

The set $\mathscr{C}(\theta)$ of Fisher information matrices $I(\theta)$ for all possible measurements is called the Fisher information \emph{complementarity chamber}  at $\theta$ for reasons that will become clear shortly.
If there exists a unique maximal Fisher information matrix $I_{\mrm{max}}(\theta)$, say, provided by the most informative measurement, as in the case of classical probability theory, then $\mathscr{C}(\theta)$ is represented by the intersection of two opposite cones  characterized by the  equation $0\leq I(\theta) \leq I_{\mrm{max}}(\theta)$.
Except in the one-parameter setting, however, this is generally not the case for the quantum theory (and also generalized probability theories). Additional constraints on the complementarity chamber reflect subtle information tradeoff among incompatible observables, which is a direct manifestation of the complementarity principle. Alternatively, these constraints may be understood as epistemic restrictions imposed by the underlying theory.

\subsection*{Characterize information complementarity with quantum estimation theory}
To unleash  the  potential of the ideas presented in the previous section, it is essential to understand the structure of the complementarity chamber or, equivalently, the constraints on the set of realizable Fisher information matrices. In the case of  quantum theory, a powerful tool for this purpose is  quantum estimation theory developed over the past half century \cite{Hels76book,Hole82book,GillM00,Zhu12the} (see supplementary information).

A generalized observable or measurement is determined by a set of positive operators that sum up to the identity. Given a state $\rho(\theta)$ parametrized  by  $\theta$ and an observable $\Pi=\{\Pi_\xi\}$, the probability of outcome $\xi$ is given by the Born rule, that is, $p(\xi|\theta)=\tr\{\rho(\theta)\Pi_\xi\}$. Accordingly, the Fisher information matrix takes on the form
\begin{equation}\label{eq:FisherQ}
I_{\Pi,jk}(\theta)=\sum_\xi
\frac{1}{p(\xi|\theta)}\tr\Bigl\{\frac{\partial
\rho}{\partial \theta_j}\Pi_\xi\Bigr\}
\Bigl\{\frac{\partial
\rho}{\partial \theta_k}\Pi_\xi\Bigr\}.
\end{equation}
As mentioned previously, the inverse Fisher information matrix sets a lower bound for the MSE matrix of any unbiased estimator. However, the bound is applicable only to the specific measurement.

To understand the structure of the complementarity chamber, it is desirable to find constraints on the Fisher information matrix that is measurement independent. According to  quantum estimation theory \cite{Hels76book,Hole82book,GillM00,Zhu12the},
one important such constraint is the SLD (symmetric logarithmic derivative) bound $I(\theta)\leq J(\theta)$,
where $J(\theta)$ is the SLD quantum Fisher information matrix given by
\begin{equation}
J_{j k}(\theta)=\frac{1}{2}\tr\bigl\{\rho(L_j L_k+L_k L_j)\bigr\},
\end{equation}
and $L_j$ is the SLD associated with the parameter $\theta_j$ as determined by the equation
\begin{equation}
\frac{\partial \rho(\theta)}{\partial \theta_j}=\frac{1}{2}(\rho
L_j+L_j\rho).
\end{equation}
In the one-parameter setting, the SLD bound can be saturated by measuring the observable $L(\theta)$;
so the complementarity chamber $\mathscr{C}(\theta)$ is a line segment determined by $0\leq I(\theta)\leq J(\theta)$.
In the multiparameter setting, however, the bound
generally cannot be saturated because the SLDs associated with different parameters are incompatible.

To determine the complementarity chamber in the multiparameter setting, it is necessary  to consider additional constraints on the Fisher information matrix that take into account the information tradeoff among incompatible observables.
Such information tradeoff is best characterized  by the Gill--Massar (GM) inequality~\cite{GillM00}
\begin{equation}
\tr\{J^{-1}(\theta) I(\theta)\}\leq d-1,
\end{equation}
which is applicable to any measurement  on a $d$-level system.
To understand the significance of the GM inequality, note that the state space has dimension $d^2-1$, so the upper bound in the above equation would be $d^2-1$ instead of $d-1$ if the SLD bound can always be saturated.
The GM inequality  is useful not only to studying the complementarity chamber and compatibility problem but also to studying  multiparameter quantum estimation problems \cite{GillM00,Zhu12the}.

\subsection*{Information complementarity illustrated}
As an illustration, here we determine the complementarity chamber of the qubit in comparison with that of the probability simplex.
In the case of a qubit, the GM inequality turns out to be both necessary and sufficient for characterizing the complementarity chamber. Moreover, any Fisher information matrix saturating the GM inequality can be realized by three mutually unbiased measurements \cite{GillM00, Zhu12the} (see supplementary information).  This observation is crucial to attaining the tomographic precision limit in  experiments  \cite{HouZXL15}.

In terms of the components of the Bloch vector $\vec{s}$, the inverse quantum Fisher information matrix reads
\begin{equation}\label{eq:QFIqubit}
J^{-1}(\vec{s}) =1-\vec{s}\vec{s}.
\end{equation}
When $s=0$ and thus $J=1$, the complementarity chamber is a cone that is isomorphic to the state space of subnormalized states for the three-dimensional real Hilbert space, with its base (the set of Fisher information matrices saturating the GM inequality) corresponding to normalized states. Fisher information matrices  of von Neumann measurements (determined by antipodal points on the Bloch sphere) correspond to normalized  pure states, while those of noisy von Neumann measurements correspond  to  subnormalized pure states.  When $s\neq 0$, the complementarity chamber $\mathscr{C}(\vec{s})$ is a distorted cone.   The \emph{metric-adjusted complementarity chamber} $\tilde{\mathscr{C}}(\vec{s}):=J^{-1/2}(\vec{s})\mathscr{C}(\vec{s})J^{-1/2}(\vec{s})$, nevertheless,  has the same size and shape irrespective of the parameter point.

\begin{figure}[t]
  \centering
  \includegraphics[width=6.8cm]{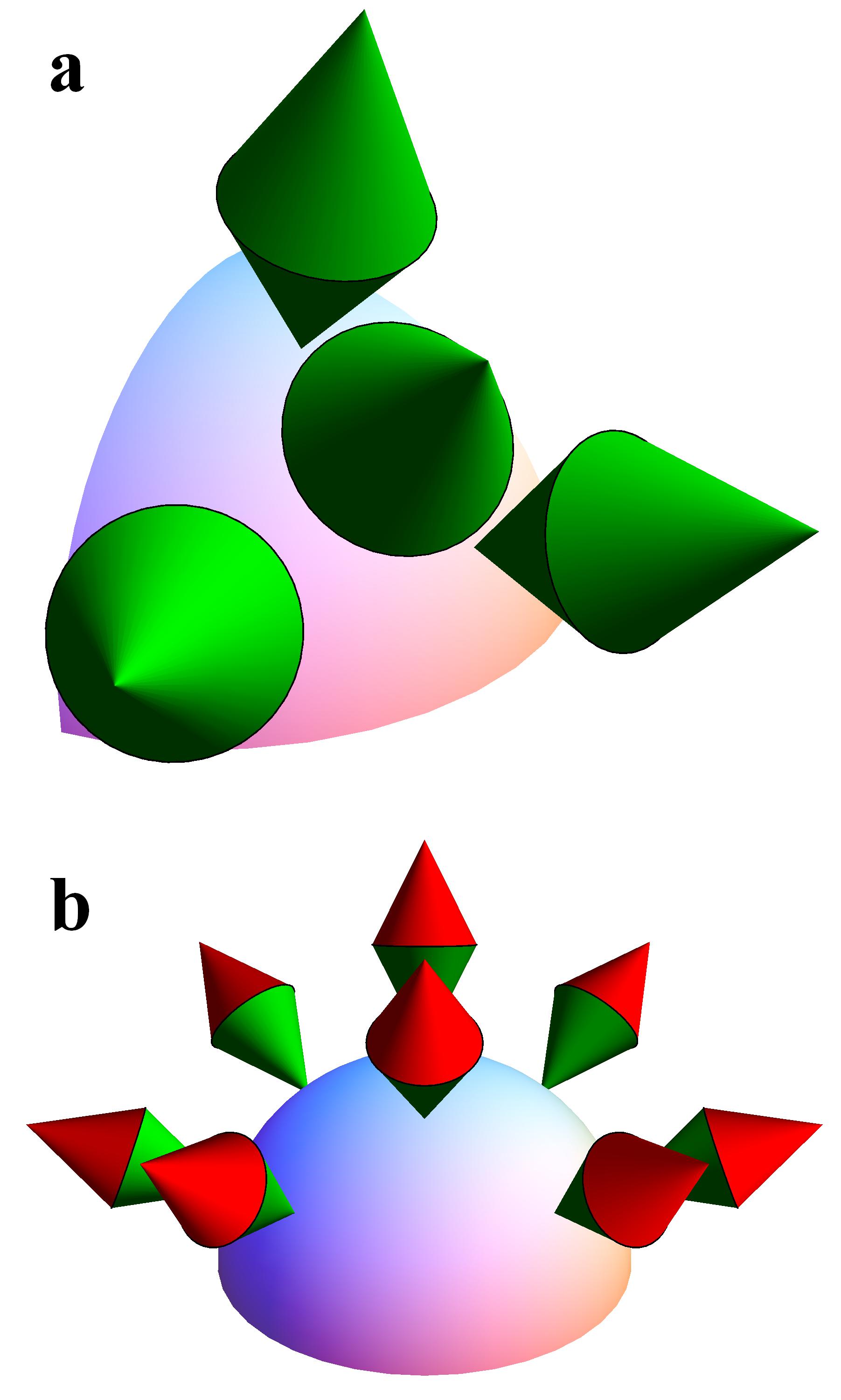}
  \caption{\label{fig:CompleChamber}\footnotesize (color online) \textbf{Metric-adjusted complementarity chambers.}  (a) Chambers (green cones, with modified size and aspect ratio for ease of viewing) on the probability simplex with respect to the Fisher--Rao metric~\cite{BengZ06book}. (b) Chambers on the state space of the real qubit with respect to the quantum Fisher information metric.   Each red cone represents the set of hypothetical Fisher information matrices satisfying the SLD bound  but excluded by the GM inequality.
   }
\end{figure}

To visualize the complementarity chamber, it is instructive to consider the real qubit. With respect to the quantum Fisher information metric  \cite{BengZ06book}, the state space is a hemisphere. Each metric-adjusted complementarity chamber is isomorphic to the state space for the two-dimensional real Hilbert space, and is represented by a circular cone, as illustrated  in the lower plot of \fref{fig:CompleChamber}. This is in sharp contrast with the complementarity chamber on the probability simplex (with three components), which is represented by the union of two opposite cones; see the upper plot of \fref{fig:CompleChamber}. The missing cone of hypothetical Fisher information matrices for the real qubit is excluded by the GM inequality.  \Fref{fig:CompleChamber} is a vivid manifestation of the viewpoint that regards quantum theory   as a classical probability theory with epistemic restrictions.

\subsection*{Universal criteria for detecting incompatible observables}
In this section we introduce a family of universal criteria for detecting incompatible observables, which are applicable to arbitrary number of arbitrary observables. So far we are not aware of any other criterion in the literature with such a wide scope of applicability. Our work fills an important gap on detecting incompatible observables and provides valuable  insight on the joint measurement problem. In addition, our incompatibility criteria can be turned into criteria for detecting EPR steering given the close connection between the two subjects \cite{UolaMG14,QuinVB14}; more details will be presented elsewhere.

Two (generalized) observables or measurements $\ob{A}=\{A_\xi\}$  and $\ob{B}=\{B_\zeta\}$ are \emph{compatible} or jointly measurable if they admit a \emph{joint observable} $\ob{M}=\{M_{\xi\zeta}\}$, which satisfies
\begin{equation}
\sum_{\zeta}M_{\xi\zeta}=A_\xi,\quad \sum_{\xi}M_{\xi\zeta}=B_\zeta.
\end{equation}
In that case, $\ob{A}$ and $\ob{B}$ are called \emph{marginal observables} of $\ob{M}$. Equivalently, $\ob{A}$  and $\ob{B}$ are compatible if they are coarse graining of a common observable $\ob{G}=\{G_\alpha\}$, that is,
\begin{equation}
A_\xi=\sum_\alpha \Lambda^{\ob{A}}_{\xi\alpha} G_{\alpha}, \quad B_\zeta=\sum_\alpha \Lambda^{\ob{B}}_{\zeta\alpha} G_{\alpha},
\end{equation}
where $\Lambda^{\ob{A}}$ and $\Lambda^{\ob{B}}$ are two stochastic matrices \cite{AliCHT09}.
Compatibility of more than two observables is defined similarly.

Suppose $\ob{M}$ is a joint observable of the set of  observables $\ob{A}_j$; then $I_{\ob{M}}(\theta)\geq I_{\ob{A}_j}(\theta)$  for any parameter point $\theta$ according to  the Fisher information data-processing inequality \cite{Zami98}. Geometrically, this inequality means that $I_{\ob{M}}(\theta)$ lies in the cone $\mathscr{V}_{\ob{A}_j}(\theta):=\{I | I\geq I_{\ob{A}_j}(\theta)\}$ of hypothetical Fisher information matrices. If the $\ob{A}_j$ are compatible, then the intersection $\cap_j\mathscr{V}_{\ob{A}_j}(\theta)$ cannot be disjoint from the complementarity chamber~$\mathscr{C}(\theta)$.
This constraint encodes  a universal criterion on the compatibility of these observables.

A simpler compatibility criterion can be derived  based on the observation that the Fisher information matrix $I_{\ob{M}}$ ($\theta$ is omitted for simplicity)  needs to satisfy the GM inequality. Define
$\tilde{I}_{\ob{A}_j}:=J^{-1/2} I_{\ob{A}_j}J^{-1/2}$ as the \emph{metric-adjusted Fisher information matrix} and
\begin{equation}\label{eq:UniversalGMT}
t(\{\tilde{I}_{\ob{A}_j}\}):=\min \{\tr \tilde{I} | \tilde{I} \geq \tilde{I}_{\ob{A}_j}\; \text{for all $j$} \}.
\end{equation}
Then $t(\{\tilde{I}_{\ob{A}_j}\})$ sets a lower bound for the GM trace $\tr(J^{-1}I_{\ob{M}})$ of any hypothetical joint observable  $\ob{M}$ of observables $\ob{A}_j$.
If the $\ob{A}_j$ are jointly measurable, then it must hold that
\begin{equation}\label{eq:UniversalCriteria}
t(\{\tilde{I}_{\ob{A}_j}\})\leq d-1,
\end{equation}
which yields  a whole family of
universal criteria for detecting incompatible observables upon varying the parameter point. These criteria are very easy to verify since $t(\{\tilde{I}_{\ob{A}_j}\})$ can be computed  with semidefinite programming. The violation of the above inequality has a clear physical interpretation: Any hypothetical joint measurement of the $\ob{A}_j$ will enable estimating certain parameters with error  at least $t(\{\tilde{I}_{\ob{A}_j}\})/(d-1)$ times smaller than allowed by the quantum theory. To see this, let $I$ be the  Fisher information matrix provided by a hypothetical joint measurement $\ob{M}$ and $\tilde{I}=J^{-1/2} I J^{-1/2}$. Then $t_{\ob{M}}:=\tr (\tilde{I}) \geq t(\{\tilde{I}_{\ob{A}_j}\})$.
Setting $W=I J^{-1} I$ as the weighting matrix, then
the GM bound for the weighted  MSE  of any unbiased estimator is given by $t_{\ob{M}}^2/(d-1)$ (see  supplementary information). By contrast, the value achievable by the hypothetical joint measurement is $\tr (W I^{-1})= t_{\ob{M}}$, which is $t_{\ob{M}}/(d-1)$ times smaller than the GM bound.

The function $t(\{\tilde{I}_{\ob{A}_j}\})$ also enjoys one of two basic requirements on a good incompatibility measure, that is, monotonicity under coarse graining (see methods section).
It is also unitarily invariant and thus may serve as a good incompatibility measure when the number of parameters under consideration is equal to the dimension $d^2-1$ of the state space and the parameter point corresponds to the completely mixed state. This incompatibility measure, denoted by $\tau(\{\ob{A}_j\})$ henceforth, can be expressed in a way that is manifestly parametrization independent and unitarily invariant (see supplementary information),
\begin{equation}\label{eq:IncompatibilityM}
\tau(\{\ob{A}_j\}):=t(\{\barcal{G}_{\ob{A}_j}\})=t(\{\mathcal{G}_{\ob{A}_j}\})-1,
\end{equation}
where $\mathcal{G}_{\ob{A}_j}$ and $\barcal{G}_{\ob{A}_j}$ are metric-adjusted  Fisher information matrices in superoperator form,
\begin{equation}\label{eq:GsuperO}
\mathcal{G}_{\ob{A}}=\sum_\xi \dket{A_\xi}\frac{1}{\tr (A_\xi)}\dbra{A_\xi},\quad  \barcal{G}_{\ob{A}}=\sum_\xi \dket{\bar{A}_\xi}\frac{1}{\tr (A_\xi)}\dbra{\bar{A}_\xi},
\end{equation}
and $\bar{A}_\xi=A_\xi-\tr(A_\xi)/d$. In the above equation, operators $A_\xi$ are taken as kets in the Hilbert-Schmidt space with the inner product $\dinner{E}{F}:=\tr(E^\dag F)$; the double ket notation is adopted to distinguish operator kets from ordinary kets in the Hilbert space \cite{Zhu12the,Zhu14IOC}. Superoperators, such as $\mathcal{G}_{\ob{A}}$, act on operator kets in the same way as operators act on ordinary kets. The threshold of the incompatibility measure $\tau(\cdot)$ is $d-1$. To reset the threshold when necessary, we may consider monotonic functions of $\tau$, such as $\max\{\tau-(d-1),0\}$ or $\max\{\tau/(d-1),1\}$.

\subsection*{Universal measurement uncertainty relations}
In this section we derive a family of universal measurement uncertainty relations, which are applicable to arbitrary number of arbitrary observables. As far as we know, no uncertainty relations with the same scope of applicability have been found before.

When a set of observables are incompatible, any approximate joint measurement entails certain degree of noisiness, which is a manifestation of  measurement uncertainty relations  \cite{MartM90,Ozaw03, BuscS06,BuscHL07,BuscLW13,BuscHOW14}. A natural way of modeling noise on an observable, say $\ob{A}=\{A_\xi\}$, is coarse graining: $\ob{A}(\Lambda):=\{A_\xi(\Lambda)=\sum_{\zeta}\Lambda_{\xi\zeta} A_\zeta\}$, where $\Lambda$ is a stochastic matrix   characterizing the noise. Of particular interest is the type of coarse graining characterized  by a single parameter: $\ob{A}(\eta)=\{\eta A_\xi+(1-\eta)\tr(A_\xi)/d\}$ with $0\leq \eta\leq 1$. Coarse graining usually reduces the information gain;  for example, $I_{\ob{A(\eta)}}=\eta^2 I_{\ob{A}}$ according to \eref{eq:FisherQ}.

Suppose $\ob{A}_j(\Lambda_j)$ is a coarse graining of
 the observable $\ob{A}_j$ characterized by the stochastic matrix $\Lambda_j$. \Eref{eq:UniversalCriteria} applied to the $\ob{A}_j(\Lambda_j)$ yields a family of  universal uncertainty relations on the strengths of measurement noises,
\begin{equation}\label{eq:MeasUncertainty}
t(\{\tilde{I}_{\ob{A}_j(\Lambda_j)}\})\leq d-1.
\end{equation}
This equation means that to measure a set of incompatible observables approximately, individual observables must be noisy enough, so  that they do not provide too much information than allowed by quantum mechanics.
As far as we know, these measurement uncertainty relations are the only known examples that are   applicable to arbitrary number of arbitrary observables.
A special but important instance of \eref{eq:MeasUncertainty} takes on the form
\begin{equation}\label{eq:MeasUncertainty2}
\tau(\{\ob{A}_j(\Lambda_j)\})=t(\{\barcal{G}_{\ob{A}_j(\Lambda_j)}\})\leq d-1,
\end{equation}
which is obtained when the number of parameters is equal to the dimension of the state space, and the parameter point corresponds to the completely mixed state.
This equation reduces to $t(\{\eta_j^2\barcal{G}_{\ob{A}_j}\})\leq d-1$ when the noise on each observable $\ob{A}_j$ is characterized by a single parameter $\eta_j$, given that $\barcal{G}_{\ob{A}_j(\eta_j)}=\eta_j^2\barcal{G}_{\ob{A}_j}$.
If in addition all $\eta_j$ are equal to $\eta$, then we have $\tau(\{\ob{A}_j(\eta)\})=\eta^2\tau(\{\ob{A}_j\})$, which leads to a simple measurement uncertainty relation,
\begin{equation}\label{eq:MeasUncertainty3}
\eta^2\leq  \frac{d-1}{\tau(\{\ob{A}_j\})}.
\end{equation}
The incompatibility measure $\tau(\{\ob{A}_j\})$ sets a lower bound for the amount of noise necessary for implementing an approximate joint measurement.

\subsection*{Coexistence of qubit effects}
To illustrate the power of our approach, here we  consider the joint measurement problem of two noisy von Neumann observables $\ob{A}=\{A,1-A\}$ and $\ob{B}=\{B,1-B\}$ in the case of a qubit, where $A=(1+\vec{a}\cdot\vec{\sigma})/2$ and $B=(1+\vec{b}\cdot\vec{\sigma})/2$. This problem is equivalent to the coexistence problem of the two effects $A$ and $B$, which has attracted substantial  attention recently \cite{Busc86,BuscS06,StanRH08,BuscS10, YuLLO10}. Most known approaches rely on mathematical tricks  tailored to this special scenario and allow no generalization. By contrast, our solution follows from a universal recipe based on simple information theoretic ideas.

\begin{figure}
  \centering
\includegraphics[width=8cm]{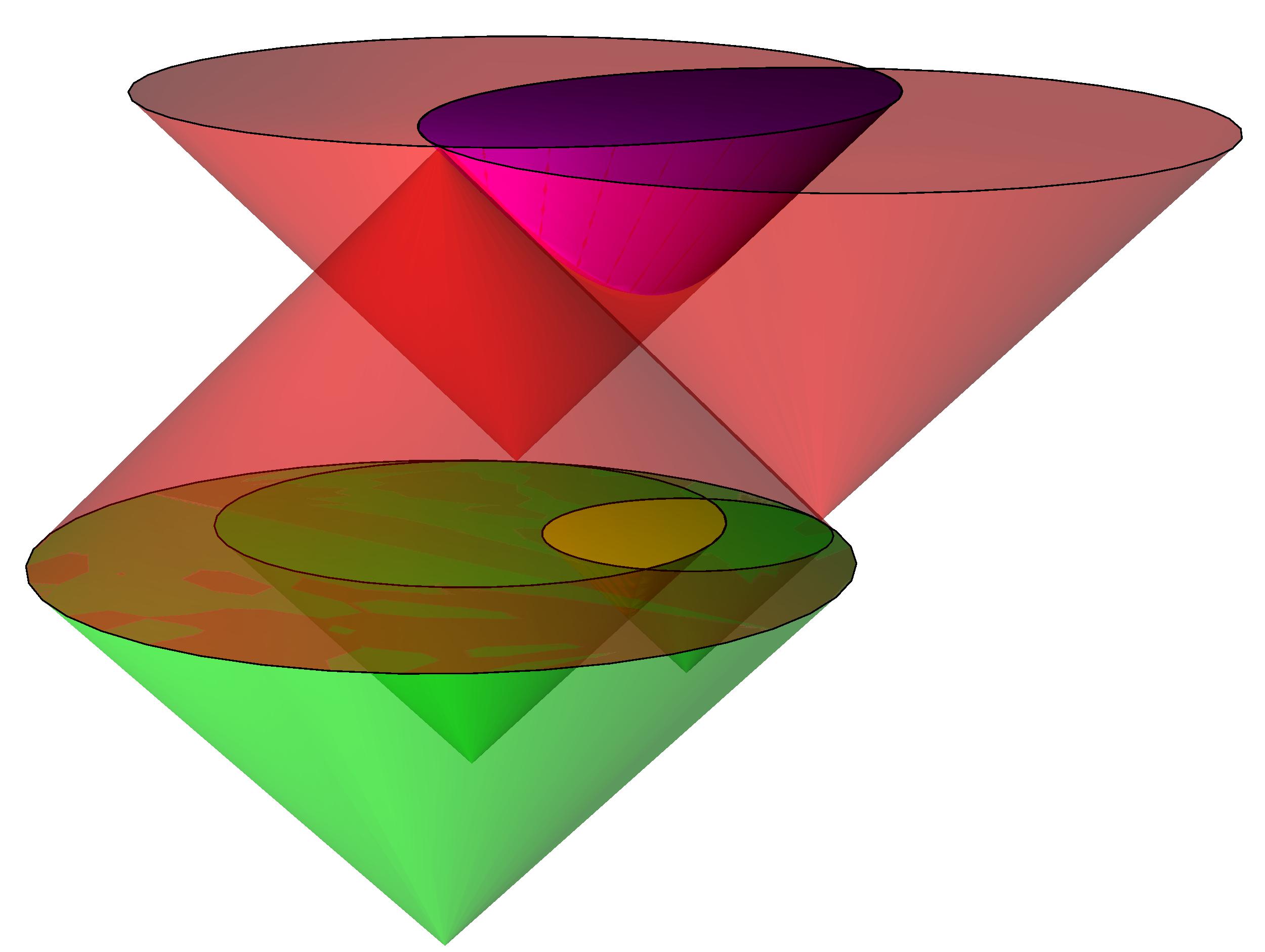}
  \caption{\label{fig:InfGeo}\footnotesize (color online) \textbf{Information geometry of qubit observables.} The largest green cone represents the complementarity chamber at the completely mixed state (cf. \fref{fig:CompleChamber}). The two upward red cones represent the sets of hypothetical Fisher information matrices lower bounded by the Fisher information matrices of two sharp von Neumann observables (corresponding to the tips  of the cones), respectively. The two observables are incompatible since the intersection of the two cones is disjoint from the complementarity chamber. The distance from the intersection to the base of the complementarity chamber quantifies the degree of incompatibility. By contrast, their noisy versions corresponding to the tips of the two smaller green cones are compatible.}
\end{figure}

According to \eref{eq:QFIqubit}, when  $s=0$, the  quantum Fisher information matrix is equal to the identity.
The Fisher information matrices of the two observables $\ob{A}$ and $\ob{B}$ are $I_{\ob{A}}=\vec{a}\vec{a}$ and $I_{\ob{B}}=\vec{b}\vec{b}$, respectively. Consequently,
\begin{align}\label{eq:tauQubit}
\tau(\ob{A},\ob{B})=\frac{1}{2}\bigl[a^2+b^2+\sqrt{(a^2+b^2)^2-4(\vec{a}\cdot\vec{b})^2}\,\bigr].
\end{align}
Remarkably, the inequality $\tau(\ob{A},\ob{B})\leq 1$ turns out to be both necessary and sufficient for the coexistence of $A$ and $B$. To verify this claim, it suffices to show its equivalence to the inequality $\norm{\vec{a}+\vec{b}}+\norm{\vec{a}-\vec{b}}\leq 2$  derived by Busch~\cite{Busc86}, which is known to be both necessary and sufficient.
Here the incompatibility measure $\tau(\ob{A},\ob{B})$ has a simple geometrical interpretation as the height (up to a scale) of
the intersection  $\mathscr{V}_{\ob{A}}\cap \mathscr{V}_{\ob{B}}$ of two cones  from the tip of the complementarity chamber, as illustrated in \fref{fig:InfGeo} (actually, this observation also offers a simple recipe for deriving  $\tau(\ob{A},\ob{B})$). The inequality $\tau(\ob{A},\ob{B})\leq 1$ means that the intersection  is not disjoint from the complementarity chamber. Otherwise, $\tau(\ob{A},\ob{B})-1$ represents the distance from the intersection to the base of the  chamber.

\subsection*{Incompatibility of noncommuting  sharp observables} It is well known that  sharp observables are compatible if and only if they commute \cite{Neum55}. However, most known proofs  rely on  mathematical tricks without physical intuition. Here
we reveal a simple information theoretic argument.

Commuting sharp observables are obviously compatible. To prove the converse, that is, compatible sharp observables commute with each other, it suffices to show that any observable $\ob{A}$ that refines a sharp observable $\ob{P}$ commutes with it. According to the Fisher information data-processing inequality \cite{Zami98}, $I_{\ob{P}}(\theta)\leq I_{\ob{A}}(\theta)$, which implies that $\barcal{G}_{\ob{P}} \leq \barcal{G}_{\ob{A}}$ and $\mathcal{G}_{\ob{P}} \leq \mathcal{G}_{\ob{A}}$ according to Sec. 3 in supplementary information, that is
\begin{equation}
\sum_\xi\frac{\douter{P_\xi}{P_\xi}}{\tr P_\xi}\leq\sum_\xi \frac{\douter{A_\xi}{A_\xi}}{\tr A_\xi}.
\end{equation}
Taking inner product with $\douter{P_\zeta}{P_\zeta}$ yields
\begin{equation}
r_\zeta\leq  \sum_\xi \frac{\dinner{P_\zeta}{A_\xi}\dinner{A_\xi}{P_\zeta}}{\tr A_\xi}\leq \sum_\xi \tr(A_\xi P_\zeta) =r_\zeta,
\end{equation}
 where $r_\zeta$ is the rank of $P_\zeta$.
The inequalities are saturated if and only if each $A_\xi$ is supported either on the range of $P_\zeta$ or on its orthogonal complement.  Therefore, $\ob{A}$ commutes with $\ob{P}$.

 The degree of incompatibility of  von Neumann observables (nondegenerate sharp observables) can be quantified by the  measure $\tau$, which turns out to be faithful now. Consider two such observables $\ob{A}$ and $\ob{B}$,  observing that $\barcal{G}_{\ob{A}}$ and   $\barcal{G}_{\ob{B}}$ are rank-($d-1$) projectors, we have
\begin{equation}
\tau(\ob{A},\ob{B})=\sum_{j=1}^{d-1}\bigl(1+\sqrt{1-s_j^2}\,\bigr),
\end{equation}
where the $s_j$ are singular values of $\barcal{G}_{\ob{A}}\barcal{G}_{\ob{B}}$  in decreasing order. The minimum $d-1$ of $\tau(\ob{A},\ob{B})$ is attained when the first $d-1$ singular values are all equal to 1, which amounts to the requirement $\mathcal{G}_{\ob{A}}=\mathcal{G}_{\ob{B}}$, that is,  $\ob{A}=\ob{B}$. The maximum $2(d-1)$  is attained when all the singular values  vanish, which happens if and only if $\ob{A}$ and $\ob{B}$ are complementary \cite{DurtEBZ10}.

Our approach  also provides a
 universal measurement uncertainty relation between $\ob{A}$ and $\ob{B}$ as characterized by the inequality $\tau(\ob{A}(\lambda),\ob{B}(\mu))\leq d-1$, where
\begin{equation}
\tau(\ob{A}(\lambda),\ob{B}(\mu))=\sum_{j=1}^{d-1}\frac{\lambda^2+\mu^2+\sqrt{(\lambda^2+\mu^2)^2-4\lambda^2\mu^2 s_j^2}}{2}.
\end{equation}
This inequality succinctly summarizes the information tradeoff between two von Neumann observables.

\subsection*{Complementary observables and quantitative wave--particle duality}
In this section we provide a simple information theoretic explanation of quantitative wave--particle duality and derive several  complementary relations that are applicable to arbitrary number of complementary observables.
Our study provides a natural framework for generalizing  previous works specializing in  the information tradeoff between two complementary observables associated with path information and fringe visibility, respectively  \cite{WootZ79, JaegSV95, Engl96}.

The complementarity
principle states that quantum
systems possess properties that are equally real but mutually
exclusive \cite{Bohr28, WootZ79, SculEW91,JaegSV95,Engl96}.  In the quintessential example of
the double-slit experiment, the photons (or electrons) may exhibit
either particle behavior or wave behavior, but the sharpening of the
particle behavior is necessarily accompanied by the blurring of
the wave behavior, and vice versa.
This wave--particle dual behavior is a manifestation of the impossibility of simultaneously measuring  complementary observables \cite{Hall95,ColeKW14}, say $\sigma_x$ and $\sigma_z$. Any attempt to acquire information about both observables is restricted by certain measurement uncertainty relation.  For example, the two unsharp observables $\ob{A}=\{(1\pm \eta_x\sigma_x)/2\}$  and $\ob{B}=\{(1\pm \eta_z\sigma_z)/2\}$ are jointly measurable if and only if \cite{Busc86,BuscS06}
\begin{equation}\label{eq:etaXZ}
\eta_x^2+\eta_z^2\leq 1.
\end{equation}
Coincidentally, this inequality is an immediate consequence of our general inequality $\tau(\ob{A},\ob{B})\leq 1$ inspired by simple information theoretic ideas. Therefore, wave--particle duality can be understood as  an epistemic restriction on the information content of observation.

Complementary relations, however, are not restricted to  two observables. The potential of our approach lies in its capability in dealing with arbitrary number of observables.
Suppose  $\ob{A}_j(\eta_j)$ are unsharp versions of complementary observables $\ob{A}_j=\{\ob{A}_{j\xi}\}$. Then $\barcal{G}_{\ob{A}_j(\eta_j)}=\eta_j^2 \barcal{G}_{\ob{A}_j}$, where $\barcal{G}_{\ob{A}_j(\eta_j)}$ and $\barcal{G}_{\ob{A}_j}$
are metric-adjusted Fisher information superoperators.
In addition,  $\barcal{G}_{\ob{A}_j}$ are mutually orthogonal rank-($d-1$) projectors. Therefore,
\begin{equation}\label{eq:tauComple}
\tau(\{\ob{A}_j(\eta_j)\}) =t(\{\eta_j^2 \barcal{G}_{\ob{A}_j}\}) =(d-1) \sum_j\eta_j^2.
\end{equation}
If the  $\ob{A}_j(\eta_j)$ are jointly measurable, then the inequality $\tau(\{\ob{A}_j\})\leq d-1$ generalizes \eref{eq:etaXZ} by setting a universal bound for the degree of unsharpness of these observables,
\begin{equation}\
 \sum_j\eta_j^2\leq 1.
\end{equation}

More generally, if the unsharpness of each observable $\ob{A}_j$ is characterized by a doubly stochastic matrix $\Lambda_j$, then \eref{eq:tauComple}  generalizes to
\begin{equation}
\tau(\{\ob{A}_j(\Lambda_j)\})=\sum_j \tr\Bigl(\Lambda_j-\frac{1}{d} K\Bigr)^2,
\end{equation}
where $K$ is the matrix with all entries equal to 1. Again,  the inequality $\tau(\{\ob{A}_j(\Lambda_j)\})\leq d-1$ constrains the information tradeoff among complementary
 observables $\ob{A}_j$.

\subsection*{Bell inequality}
Our simple information theoretic ideas can also shed new light on Bell nonlocality~\cite{Bell64, BrunCPS13}. As an illustration, here we show that given two observables for one party, the maximum violation of the CHSH inequality~\cite{ClauHSH69} is a simple function of the measure of incompatibility introduced in this paper. Since Bell nonlocality may be seen as a special instance of contextuality~\cite{KochS67,LianSW11,CabeSW14}, our work is also helpful  to this latter subject.

Suppose  we have  two $\pm 1$ valued observables  $A$  and $B$ for party~1 together with similar observables $C$ and $D$ for party~2 (here we use Hermitian operators to represent observables following common convention; $A$ is equivalent to $\ob{A}=\{A_\pm\}$ in our convention, where $A_{\pm}$ are the eigenprojectors of $A$). A bipartite state $\rho$ satisfies the CHSH inequality if and only if $|\langle \mathbb{B}\rangle_\rho|\leq 1$~\cite{Land87,WolfPF09}, where
\begin{equation}
\mathbb{B}=\frac{1}{2}[A\otimes (C+D)+B\otimes (C-D)]
\end{equation}
is the CHSH operator and satisfies
\begin{equation}
\mathbb{B}^2=1+\frac{1}{4}[A,B]\otimes [C,D].
\end{equation}
Given the observables $A$ and $B$ for party~1, the maximal violation of the CHSH inequality is attained when $C$ and $D$ are anticommuting Pauli matrices \cite{Land87,WolfPF09},
\begin{equation}
\max_{\rho, C,D} |\langle \mathbb{B}\rangle_\rho|=\sqrt{1+\frac{1}{2} \norm{[A,B]}}.
\end{equation}
In the case party~1 is a qubit, suppose $A=\vec{a}\cdot \vec{\sigma}$ and $B=\vec{b}\cdot \vec{\sigma}$ with unit vectors $\vec{a}$ and $\vec{b}$. Then
\begin{equation}
\max_{\rho, C,D} |\langle \mathbb{B}\rangle_\rho|=\sqrt{1+\sin\theta}=\sqrt{\tau(A,B)},
\end{equation}
where $\theta$ is the angle spanned by  vectors $\vec{a}$ and $\vec{b}$. Remarkably,  the  maximum is  equal to the square root of the measure of incompatibility of $A$ and $B$ built on simple information theoretic ideas. This observation may have profound implications for understanding Bell inequalities from information theoretic perspectives.

 In general, we can find spectral decompositions  $A_+=\sum_j \outer{\psi_j}{\psi_j}$ and $B_+=\sum_k \outer{\varphi_k}{\varphi_k}$ (which correspond to the singular value decomposition of $A_+B_+$) such that $\inner{\psi_j}{\varphi_k}=\delta_{jk}\cos(\theta_j /2)$ with $0\leq \theta_j\leq \pi$. Without loss of generality, we assume $\max_j \sin\theta_j=\sin\theta_1>0$. Then
\begin{equation}
\max_{\rho, C,D} |\langle \mathbb{B}\rangle_\rho|=\sqrt{1+\sin \theta_1}=\sqrt{\tau(A^\prime,B^\prime)},
\end{equation}
and the maximum is attained at a Bell state whose local support for party~1 is spanned by $|\psi_1\rangle$ and $|\phi_1\rangle$. Here $A^\prime$ and $B^\prime$ are the restrictions of $A$ and $B$ on this two-dimensional subspace.

\section*{Summary}  We have introduced a new paradigm for detecting and characterizing  incompatible observables starting from two simple information theoretic ideas, quite in the spirit of the slogan "physics is informational".  Unlike most studies on this subject based on Shannon information, our work employs  Fisher information to capture the information tradeoff among incompatible observables, which turns out to be surprisingly effective. This line of thinking is quite   fruitful in studying a number of  foundational issues.
In particular, we introduced a family of universal criteria for detecting incompatible observables, which are applicable to arbitrary number of arbitrary observables.  These criteria   fill an important gap on detecting incompatible observables and provide valuable  insight on the joint measurement problem. They are also useful for detecting EPR steering given the close connection between steering and incompatible observables. The same idea also leads to a natural measure of incompatibility, which can easily be computed by semidefinite programming.   By virtue of this framework, we derived a family of universal  measurement uncertainty relations, which are applicable to arbitrary number of arbitrary observables. In addition, our work provided a simple information theoretic explanation of quantitative wave--particle duality and offered new perspectives for understanding Bell nonlocality, contextuality, and quantum precision limit. Our study is of interest to researchers from diverse fields, such as information theory, quantum estimation theory, quantum metrology, and quantum foundations.

\bigskip

\section*{Methods}
\subsection*{Measures of incompatibility}
 Here we discuss briefly how to quantify the degree of incompatibility of a set of observables, motivated by the discussions in the main text.  To simplify the notation, we  focus on two observables, say $\ob{A}=\{A_\xi\}$ and $\ob{B}=\{B_\zeta\}$; the generalization to more observables is immediate.

\subsubsection*{Basic requirements}
Like  an entanglement measure, any good incompatibility measure, say $\tau(\ob{A}, \ob{B})$, should satisfy certain basic requirements, among which the following two are very natural:
\begin{enumerate}
\item Unitary invariance: $\tau(U \ob{A} U^\dag, U \ob{B} U^\dag )=\tau(\ob{A}, \ob{B})$;
\item Monotonicity  under coarse graining.
\end{enumerate}
Additional requirements, such as continuity, faithfulness, and choices of the scale and threshold may be imposed if necessary. To ensure great generality, however, we shall retain only the most basic requirements. Here the first requirement is self explaining. To make the second one more precise, we need to introduce an order relation on observables following Martens and de Muynck~\cite{MartM90}.

Observable $\ob{C}$ is a \emph{coarse graining} of $\ob{A}$ if
  $C_\xi=\sum_\zeta\Lambda_{\xi\zeta} A_\zeta$  for some stochastic matrix $\Lambda_{\xi\zeta}$, which satisfies $\Lambda_{\xi\zeta}\geq 0$ and $\sum_{\xi} \Lambda_{\xi\zeta}=1$. By contrast, we say $\ob{A}$ is a \emph{refinement} of $\ob{C}$. This order relation is  denoted by $\ob{C}\overset{\Lambda}{\preceq} \ob{A}$ (or $\ob{A}\overset{\Lambda}{\succeq} \ob{C}$), where the symbol $\Lambda$ may be omitted if it is of no concern.
It has a clear operational interpretation: Any setup that realizes the observable $\ob{A}$ can also realize $\ob{C}$ with suitable data processing as specified by the stochastic matrix. It is straightforward to verify that the order relation just defined is reflexive and transitive. Two observables $\ob{A}$ and $\ob{C}$ are \emph{equivalent} if $\ob{C}\preceq \ob{A}$ and  $\ob{A}\preceq \ob{C}$. Such observables provide the same amount of information and may be identified if we are only concerned with their information contents. The resulting order relation  on equivalent classes is antisymmetric in addition to being reflexive and transitive, and is thus a partial order.

Suppose four observables $\ob{A},\ob{B},\ob{C},\ob{D}$ satisfy $\ob{C}\preceq \ob{A}$ and $\ob{D}\preceq \ob{B}$. If $\ob{A}$ and $\ob{B}$ are compatible, then $\ob{C}$ and $\ob{D}$ are also compatible.  Requirement~2  on the incompatibility measure amounts to the inequality $\tau(\ob{C}, \ob{D})\leq \tau(\ob{A}, \ob{B})$, which may be seen as a natural extension of the above intuition.

\subsubsection*{Robustness}
A simple incompatibility measure can be defined in analogy with the entanglement measure  robustness. Define $\ob{A}_\epsilon=\{A_{\xi\epsilon}\}$ with
\begin{equation}
A_{\xi\epsilon}=\frac{A_\xi+\frac{\epsilon \tr(A_\xi)} {d }}{1+\epsilon},
\end{equation}
 The \emph{robustness} $R(\ob{A},\ob{B})$ of two observables $\ob{A}$ and $\ob{B}$  is defined as the minimal nonnegative number $\epsilon$ such that $\ob{A}_\epsilon$ and $\ob{B}_\epsilon$ are compatible. A close relative of this measure is the logarithmic robustness  $R_{\mrm{L}}(\ob{A},\ob{B}):=\ln [1+R(\ob{A},\ob{B})]$. It is straightforward to verify that the robustness is unitarily invariant and faithful. To show monotonicity under coarse graining, note that  $\ob{C}_\epsilon\overset{\Lambda}{\preceq} \ob{A}_\epsilon$ whenever $\ob{C}\overset{\Lambda}{\preceq} \ob{A}$. Suppose $\ob{C}\preceq \ob{A}$ and $\ob{D}\preceq \ob{B}$; then $\ob{C}_\epsilon$ and $\ob{D}_\epsilon$ are compatible whenever  $\ob{A}_\epsilon$ and $\ob{B}_\epsilon$ are. So $R(\ob{C},\ob{D})\leq R(\ob{A};\ob{B})$; that is, the robustness is nonincreasing under coarse graining.

\subsubsection*{Incompatibility measure inspired by quantum estimation theory}
In  this section, we introduce an incompatibility measure based on quantum estimation theory and  simple information theoretic ideas presented in the main text. It is easy to compute and is useful for detecting and characterizing incompatible observables.

Our starting point is the observation that $I_\ob{C}(\theta)\leq I_\ob{A}(\theta)$ whenever $\ob{C}\preceq \ob{A}$, as follows from the Fisher information data-processing inequality~\cite{Zami98}.  In particular, the Fisher information has the nice property of being independent of representative observables in a given equivalent class. For example, it is invariant under relabeling of outcomes or "splitting" of an outcome, say $A_\xi\rightarrow \{A_\xi/2, A_\xi/2\}$, which has little physical significance. We note that few other information or uncertainty measures satisfy this natural requirement.

As an implication of the above analysis, $t(\tilde{I}_\ob{A}(\theta), \tilde{I}_\ob{B}(\theta))$ is monotonic under coarse graining,  where $\tilde{I}_{\ob{A}}=J^{-1/2} I_{\ob{A}}J^{-1/2}$ is the metric-adjusted Fisher information and $t(\tilde{I}_\ob{A}(\theta), \tilde{I}_\ob{B}(\theta))$ is defined in \eref{eq:UniversalGMT} in the main text. If the number of parameters is equal to $d^2-1$, and the parameter point $\theta$ corresponds to the completely mixed state, then
\begin{equation}
t(\tilde{I}_\ob{A}(\theta), \tilde{I}_\ob{B}(\theta))=t(\barcal{G}_{\ob{A}}, \barcal{G}_\ob{B})=t(\mathcal{G}_{\ob{A}}, \mathcal{G}_\ob{B})-1
\end{equation}
according to Sec. 3 in supplementary information, where $\mathcal{G}$ and $\barcal{G}$ are metric-adjusted Fisher information superoperators as specified  in \eref{eq:FrameSO-G} there and  \eref{eq:GsuperO} in the main text.
Define
\begin{equation}
\tau(\ob{A}, \ob{B}):=t(\barcal{G}_{\ob{A}}, \barcal{G}_\ob{B}),
\end{equation}
then $\tau(\ob{A}, \ob{B})$ is both unitarily invariant and monotonic, thereby satisfying the two basic requirements of a good incompatibility measure. The threshold of $\tau(\ob{A}, \ob{B})$ is $d-1$. If $\tau(\ob{A}, \ob{B})>d-1$, then  the two observables $\ob{A}$ and $\ob{B}$ are necessarily incompatible; otherwise, either possibility may happen.
To derive a measure with a usual threshold, we may opt for
$\max\{\tau(\ob{A}, \ob{B})-(d-1),0\}$
 instead of $\tau(\ob{A}, \ob{B})$. In this paper, however, we are mostly concerned with the ratio $\tau(\ob{A}, \ob{B})/(d-1)$.

Although $\tau(\ob{A}, \ob{B})$ is generally not faithful, it  provides lower bounds for  the faithful  measures $R(\ob{A},\ob{B})$ and  $R_{\mrm{L}}(\ob{A},\ob{B})$,
 \begin{align}
 R(\ob{A},\ob{B})\geq\sqrt{ \frac{\tau(\ob{A}, \ob{B})}{d-1}}-1,\quad
 R_{\mrm{L}}(\ob{A},\ob{B})\geq \frac{1}{2}\ln \frac{\tau(\ob{A}, \ob{B})}{d-1}.
 \end{align}
This equation is an immediate consequence of the observation $\tau(\ob{A}_\epsilon, \ob{B}_\epsilon)=\tau(\ob{A}, \ob{B})/(1+\epsilon)^2$.
In addition,
 it is faithful on von Neumann observables, as demonstrated in the main text.
These nice properties corroborate $\tau$ as a good incompatibility measure.


\begin{thebibliography}{10}
\expandafter\ifx\csname url\endcsname\relax
  \def\url#1{\texttt{#1}}\fi
\expandafter\ifx\csname urlprefix\endcsname\relax\def\urlprefix{URL }\fi
\providecommand{\bibinfo}[2]{#2}
\providecommand{\eprint}[2][]{\url{#2}}

\bibitem{Heis27}
\bibinfo{author}{Heisenberg, W.}
\newblock \bibinfo{title}{{\"{U}}ber den anschaulichen {Inhalt} der
  quantentheoretischen {Kinematik} und {Mechanik}}.
\newblock \emph{\bibinfo{journal}{Zeit. f\"ur Physik}}
  \textbf{\bibinfo{volume}{43}}, \bibinfo{pages}{172} (\bibinfo{year}{1927}).

\bibitem{BuscHL07}
\bibinfo{author}{Busch, P.}, \bibinfo{author}{Heinonen, T.} \&
  \bibinfo{author}{Lahti, P.}
\newblock \bibinfo{title}{\uppercase{H}eisenberg's uncertainty principle}.
\newblock \emph{\bibinfo{journal}{Phys. Rep.}} \textbf{\bibinfo{volume}{452}},
  \bibinfo{pages}{155--176} (\bibinfo{year}{2007}).

\bibitem{WehnW10}
\bibinfo{author}{Wehner, S.} \& \bibinfo{author}{Winter, A.}
\newblock \bibinfo{title}{Entropic uncertainty relations---a survey}.
\newblock \emph{\bibinfo{journal}{New J. Phys.}} \textbf{\bibinfo{volume}{12}},
  \bibinfo{pages}{025009} (\bibinfo{year}{2010}).

\bibitem{Bohr28}
\bibinfo{author}{Bohr, N.}
\newblock \bibinfo{title}{The quantum postulate and the recent development of
  atomic theory}.
\newblock \emph{\bibinfo{journal}{Nature}} \textbf{\bibinfo{volume}{121}},
  \bibinfo{pages}{580--590} (\bibinfo{year}{1928}).

\bibitem{WootZ79}
\bibinfo{author}{Wootters, W.~K.} \& \bibinfo{author}{Zurek, W.~H.}
\newblock \bibinfo{title}{Complementarity in the double-slit experiment:
  Quantum nonseparability and a quantitative statement of \uppercase{B}ohr's
  principle}.
\newblock \emph{\bibinfo{journal}{Phys. Rev. D}} \textbf{\bibinfo{volume}{19}},
  \bibinfo{pages}{473--484} (\bibinfo{year}{1979}).

\bibitem{SculEW91}
\bibinfo{author}{Scully, M.~O.}, \bibinfo{author}{Englert, B.-G.} \&
  \bibinfo{author}{Walther, H.}
\newblock \bibinfo{title}{Quantum optical tests of complementarity}.
\newblock \emph{\bibinfo{journal}{Nature}} \textbf{\bibinfo{volume}{351}},
  \bibinfo{pages}{111--116} (\bibinfo{year}{1991}).

\bibitem{JaegSV95}
\bibinfo{author}{Jaeger, G.}, \bibinfo{author}{Shimony, A.} \&
  \bibinfo{author}{Vaidman, L.}
\newblock \bibinfo{title}{Two interferometric complementarities}.
\newblock \emph{\bibinfo{journal}{Phys. Rev. A}} \textbf{\bibinfo{volume}{51}},
  \bibinfo{pages}{54--67} (\bibinfo{year}{1995}).

\bibitem{Engl96}
\bibinfo{author}{Englert, B.-G.}
\newblock \bibinfo{title}{Fringe visibility and which-way information: An
  inequality}.
\newblock \emph{\bibinfo{journal}{Phys. Rev. Lett.}}
  \textbf{\bibinfo{volume}{77}}, \bibinfo{pages}{2154--2157}
  (\bibinfo{year}{1996}).

\bibitem{BuscS06}
\bibinfo{author}{Busch, P.} \& \bibinfo{author}{Shilladay, C.}
\newblock \bibinfo{title}{Complementarity and uncertainty in {Mach}-{Zehnder}
  interferometry and beyond}.
\newblock \emph{\bibinfo{journal}{Phys. Rep.}} \textbf{\bibinfo{volume}{435}},
  \bibinfo{pages}{1--31} (\bibinfo{year}{2006}).

\bibitem{Bell64}
\bibinfo{author}{Bell, J.~S.}
\newblock \bibinfo{title}{On the {Einstein} {Podolsky} {Rosen} paradox}.
\newblock \emph{\bibinfo{journal}{Physics}} \textbf{\bibinfo{volume}{1}},
  \bibinfo{pages}{195--200} (\bibinfo{year}{1964}).

\bibitem{ClauHSH69}
\bibinfo{author}{Clauser, J.~F.}, \bibinfo{author}{Horne, M.~A.},
  \bibinfo{author}{Shimony, A.} \& \bibinfo{author}{Holt, R.~A.}
\newblock \bibinfo{title}{Proposed experiment to test local hidden-variable
  theories}.
\newblock \emph{\bibinfo{journal}{Phys. Rev. Lett.}}
  \textbf{\bibinfo{volume}{23}}, \bibinfo{pages}{880--884}
  (\bibinfo{year}{1969}).

\bibitem{BrunCPS13}
\bibinfo{author}{Brunner, N.}, \bibinfo{author}{Cavalcanti, D.},
  \bibinfo{author}{Pironio, S.}, \bibinfo{author}{Scarani, V.} \&
  \bibinfo{author}{Wehner, S.}
\newblock \bibinfo{title}{{Bell} nonlocality}.
\newblock \emph{\bibinfo{journal}{Rev. Mod. Phys.}}
  \textbf{\bibinfo{volume}{86}}, \bibinfo{pages}{419--478}
  (\bibinfo{year}{2014}).

\bibitem{WiseJD07}
\bibinfo{author}{Wiseman, H.~M.}, \bibinfo{author}{Jones, S.~J.} \&
  \bibinfo{author}{Doherty, A.~C.}
\newblock \bibinfo{title}{Steering, entanglement, nonlocality, and the
  {Einstein-Podolsky-Rosen} paradox}.
\newblock \emph{\bibinfo{journal}{Phys. Rev. Lett.}}
  \textbf{\bibinfo{volume}{98}}, \bibinfo{pages}{140402}
  (\bibinfo{year}{2007}).

\bibitem{UolaMG14}
\bibinfo{author}{Uola, R.}, \bibinfo{author}{Moroder, T.} \&
  \bibinfo{author}{G\"uhne, O.}
\newblock \bibinfo{title}{Joint measurability of generalized measurements
  implies classicality}.
\newblock \emph{\bibinfo{journal}{Phys. Rev. Lett.}}
  \textbf{\bibinfo{volume}{113}}, \bibinfo{pages}{160403}
  (\bibinfo{year}{2014}).

\bibitem{QuinVB14}
\bibinfo{author}{Quintino, M.~T.}, \bibinfo{author}{V\'ertesi, T.} \&
  \bibinfo{author}{Brunner, N.}
\newblock \bibinfo{title}{Joint measurability, {Einstein-Podolsky-Rosen}
  steering, and {Bell} nonlocality}.
\newblock \emph{\bibinfo{journal}{Phys. Rev. Lett.}}
  \textbf{\bibinfo{volume}{113}}, \bibinfo{pages}{160402}
  (\bibinfo{year}{2014}).

\bibitem{KochS67}
\bibinfo{author}{Kochen, S.} \& \bibinfo{author}{Specker, E.~P.}
\newblock \bibinfo{title}{The problem of hidden variables in quantum
  mechanics}.
\newblock \emph{\bibinfo{journal}{J. Math. Mech.}}
  \textbf{\bibinfo{volume}{17}}, \bibinfo{pages}{59} (\bibinfo{year}{1967}).

\bibitem{LianSW11}
\bibinfo{author}{Liang, Y.-C.}, \bibinfo{author}{Spekkens, R.~W.} \&
  \bibinfo{author}{Wiseman, H.~M.}
\newblock \bibinfo{title}{Specker's parable of the overprotective seer: A road
  to contextuality, nonlocality and complementarity}.
\newblock \emph{\bibinfo{journal}{Phys. Rep.}} \textbf{\bibinfo{volume}{506}},
  \bibinfo{pages}{1--39} (\bibinfo{year}{2011}).

\bibitem{AbraB11}
\bibinfo{author}{Abramsky, S.} \& \bibinfo{author}{Brandenburger, A.}
\newblock \bibinfo{title}{The sheaf-theoretic structure of non-locality and
  contextuality}.
\newblock \emph{\bibinfo{journal}{New J. Phys.}} \textbf{\bibinfo{volume}{13}},
  \bibinfo{pages}{113036} (\bibinfo{year}{2011}).

\bibitem{CabeSW14}
\bibinfo{author}{Cabello, A.}, \bibinfo{author}{Severini, S.} \&
  \bibinfo{author}{Winter, A.}
\newblock \bibinfo{title}{Graph-theoretic approach to quantum correlations}.
\newblock \emph{\bibinfo{journal}{Phys. Rev. Lett.}}
  \textbf{\bibinfo{volume}{112}}, \bibinfo{pages}{040401}
  (\bibinfo{year}{2014}).

\bibitem{Cole13}
\bibinfo{author}{Coles, P.~J.}
\newblock \bibinfo{title}{Role of complementarity in superdense coding}.
\newblock \emph{\bibinfo{journal}{Phys. Rev. A}} \textbf{\bibinfo{volume}{88}},
  \bibinfo{pages}{062317} (\bibinfo{year}{2013}).

\bibitem{Ozaw03}
\bibinfo{author}{Ozawa, M.}
\newblock \bibinfo{title}{Universally valid reformulation of the {Heisenberg}
  uncertainty principle on noise and disturbance in measurement}.
\newblock \emph{\bibinfo{journal}{Phys. Rev. A}} \textbf{\bibinfo{volume}{67}},
  \bibinfo{pages}{042105} (\bibinfo{year}{2003}).

\bibitem{BuscLW13}
\bibinfo{author}{Busch, P.}, \bibinfo{author}{Lahti, P.} \&
  \bibinfo{author}{Werner, R.~F.}
\newblock \bibinfo{title}{Proof of {Heisenberg's} error-disturbance relation}.
\newblock \emph{\bibinfo{journal}{Phys. Rev. Lett.}}
  \textbf{\bibinfo{volume}{111}}, \bibinfo{pages}{160405}
  (\bibinfo{year}{2013}).

\bibitem{BuscHOW14}
\bibinfo{author}{Buscemi, F.}, \bibinfo{author}{Hall, M. J.~W.},
  \bibinfo{author}{Ozawa, M.} \& \bibinfo{author}{Wilde, M.~M.}
\newblock \bibinfo{title}{Noise and disturbance in quantum measurements: An
  information-theoretic approach}.
\newblock \emph{\bibinfo{journal}{Phys. Rev. Lett.}}
  \textbf{\bibinfo{volume}{112}}, \bibinfo{pages}{050401}
  (\bibinfo{year}{2014}).

\bibitem{HeinW10}
\bibinfo{author}{Heinosaari, T.} \& \bibinfo{author}{Wolf, M.~M.}
\newblock \bibinfo{title}{Nondisturbing quantum measurements}.
\newblock \emph{\bibinfo{journal}{J. Math. Phys.}}
  \textbf{\bibinfo{volume}{51}}, \bibinfo{pages}{092201}
  (\bibinfo{year}{2010}).

\bibitem{ReebRW13}
\bibinfo{author}{Reeb, D.}, \bibinfo{author}{Reitzner, D.} \&
  \bibinfo{author}{Wolf, M.~M.}
\newblock \bibinfo{title}{Coexistence does not imply joint measurability}.
\newblock \emph{\bibinfo{journal}{J. Phys. A: Math. Theor.}}
  \textbf{\bibinfo{volume}{46}}, \bibinfo{pages}{462002}
  (\bibinfo{year}{2013}).

\bibitem{Neum55}
\bibinfo{author}{{von Neumann}, J.}
\newblock \emph{\bibinfo{title}{Mathematical Foundations of Quantum Mechanics}}
  (\bibinfo{publisher}{Princeton University Press},
  \bibinfo{address}{Princeton, NJ}, \bibinfo{year}{1955}).
\newblock \bibinfo{note}{Translated from the German edition by R. T. Beyer}.

\bibitem{WolfPF09}
\bibinfo{author}{Wolf, M.~M.}, \bibinfo{author}{Perez-Garcia, D.} \&
  \bibinfo{author}{Fernandez, C.}
\newblock \bibinfo{title}{Measurements incompatible in quantum theory cannot be
  measured jointly in any other no-signaling theory}.
\newblock \emph{\bibinfo{journal}{Phys. Rev. Lett.}}
  \textbf{\bibinfo{volume}{103}}, \bibinfo{pages}{230402}
  (\bibinfo{year}{2009}).

\bibitem{Busc86}
\bibinfo{author}{Busch, P.}
\newblock \bibinfo{title}{Unsharp reality and joint measurements for spin
  observables}.
\newblock \emph{\bibinfo{journal}{Phys. Rev. D}} \textbf{\bibinfo{volume}{33}},
  \bibinfo{pages}{2253--2261} (\bibinfo{year}{1986}).

\bibitem{StanRH08}
\bibinfo{author}{Stano, P.}, \bibinfo{author}{Reitzner, D.} \&
  \bibinfo{author}{Heinosaari, T.}
\newblock \bibinfo{title}{Coexistence of qubit effects}.
\newblock \emph{\bibinfo{journal}{Phys. Rev. A}} \textbf{\bibinfo{volume}{78}},
  \bibinfo{pages}{012315} (\bibinfo{year}{2008}).

\bibitem{BuscS10}
\bibinfo{author}{Busch, P.} \& \bibinfo{author}{Schmidt, H.-J.}
\newblock \bibinfo{title}{Coexistence of qubit effects}.
\newblock \emph{\bibinfo{journal}{Quantum Inf. Process.}}
  \textbf{\bibinfo{volume}{9}}, \bibinfo{pages}{143--169}
  (\bibinfo{year}{2010}).

\bibitem{YuLLO10}
\bibinfo{author}{Yu, S.}, \bibinfo{author}{Liu, N.-L.}, \bibinfo{author}{Li,
  L.} \& \bibinfo{author}{Oh, C.~H.}
\newblock \bibinfo{title}{Joint measurement of two unsharp observables of a
  qubit}.
\newblock \emph{\bibinfo{journal}{Phys. Rev. A}} \textbf{\bibinfo{volume}{81}},
  \bibinfo{pages}{062116} (\bibinfo{year}{2010}).

\bibitem{Muyn00}
\bibinfo{author}{de~Muynck, W.~M.}
\newblock \bibinfo{title}{Preparation and measurement: Two independent sources
  of uncertainty in quantum mechanics}.
\newblock \emph{\bibinfo{journal}{Found. Phys.}} \textbf{\bibinfo{volume}{30}},
  \bibinfo{pages}{205--225} (\bibinfo{year}{2000}).

\bibitem{Robe29}
\bibinfo{author}{Robertson, H.~P.}
\newblock \bibinfo{title}{The uncertainty principle}.
\newblock \emph{\bibinfo{journal}{Phys. Rev.}} \textbf{\bibinfo{volume}{34}},
  \bibinfo{pages}{163--164} (\bibinfo{year}{1929}).

\bibitem{MartM90}
\bibinfo{author}{Martens, H.} \& \bibinfo{author}{de~Muynck, W.~M.}
\newblock \bibinfo{title}{Nonideal quantum measurements}.
\newblock \emph{\bibinfo{journal}{Found. Phys.}} \textbf{\bibinfo{volume}{20}},
  \bibinfo{pages}{255--281} (\bibinfo{year}{1990}).

\bibitem{MartM90T}
\bibinfo{author}{Martens, H.} \& \bibinfo{author}{de~Muynck, W.~M.}
\newblock \bibinfo{title}{The inaccuracy principle}.
\newblock \emph{\bibinfo{journal}{Found. Phys.}} \textbf{\bibinfo{volume}{20}},
  \bibinfo{pages}{357--380} (\bibinfo{year}{1990}).

\bibitem{BuscHSS13}
\bibinfo{author}{Busch, P.}, \bibinfo{author}{Heinosaari, T.},
  \bibinfo{author}{Schultz, J.} \& \bibinfo{author}{Stevens, N.}
\newblock \bibinfo{title}{{Comparing the degrees of incompatibility inherent in
  probabilistic physical theories}}.
\newblock \emph{\bibinfo{journal}{EPL}} \bibinfo{pages}{10002}
  (\bibinfo{year}{2013}).

\bibitem{Hels76book}
\bibinfo{author}{Helstrom, C.~W.}
\newblock \emph{\bibinfo{title}{Quantum Detection and Estimation Theory}}
  (\bibinfo{publisher}{Academic Press}, \bibinfo{address}{New York},
  \bibinfo{year}{1976}).

\bibitem{Hole82book}
\bibinfo{author}{Holevo, A.~S.}
\newblock \emph{\bibinfo{title}{Probabilistic and Statistical Aspects of
  Quantum Theory}} (\bibinfo{publisher}{North-Holland},
  \bibinfo{address}{Amsterdam}, \bibinfo{year}{1982}).

\bibitem{Hard01}
\bibinfo{author}{Hardy, L.}
\newblock \bibinfo{title}{{Quantum Theory From Five Reasonable Axioms}}
  (\bibinfo{year}{2001}).
\newblock \urlprefix\url{http://arxiv.org/abs/quant-ph/0101012}.
\newblock \bibinfo{note}{Date of access: 03/01/2001}.

\bibitem{Barr07}
\bibinfo{author}{Barrett, J.}
\newblock \bibinfo{title}{Information processing in generalized probabilistic
  theories}.
\newblock \emph{\bibinfo{journal}{Phys. Rev. A}} \textbf{\bibinfo{volume}{75}},
  \bibinfo{pages}{032304} (\bibinfo{year}{2007}).

\bibitem{Fish25}
\bibinfo{author}{Fisher, R.~A.}
\newblock \bibinfo{title}{Theory of statistical estimation}.
\newblock \emph{\bibinfo{journal}{Math. Proc. Cambr. Philos. Soc.}}
  \textbf{\bibinfo{volume}{22}}, \bibinfo{pages}{700--725}
  (\bibinfo{year}{1925}).

\bibitem{GillM00}
\bibinfo{author}{Gill, R.~D.} \& \bibinfo{author}{Massar, S.}
\newblock \bibinfo{title}{State estimation for large ensembles}.
\newblock \emph{\bibinfo{journal}{Phys. Rev. A}} \textbf{\bibinfo{volume}{61}},
  \bibinfo{pages}{042312} (\bibinfo{year}{2000}).

\bibitem{Zhu12the}
\bibinfo{author}{Zhu, H.}
\newblock \emph{\bibinfo{title}{Quantum State Estimation and Symmetric
  Informationally Complete {POM}s}}.
\newblock Ph.D. thesis, \bibinfo{school}{National University of Singapore}
  (\bibinfo{year}{2012}).
\newblock \bibinfo{note}{Available at
  \url{http://scholarbank.nus.edu.sg/bitstream/handle/10635/35247/ZhuHJthesis.pdf}.
  Date of access: 31/10/2012}.

\bibitem{HouZXL15}
\bibinfo{author}{Hou, Z.}, \bibinfo{author}{Zhu, H.}, \bibinfo{author}{Xiang,
  G.-Y.}, \bibinfo{author}{Li, C.-F.} \& \bibinfo{author}{Guo, G.-C.}
\newblock \bibinfo{title}{{Experimental verification of quantum precision limit
  in adaptive qubit state tomography}} (\bibinfo{year}{2015}).
\newblock \urlprefix\url{http://arxiv.org/abs/1503.00264}.
\newblock \bibinfo{note}{Date of access: 05/03/2015}.

\bibitem{BengZ06book}
\bibinfo{author}{Bengtsson, I.} \& \bibinfo{author}{{\.{Z}}yczkowski, K.}
\newblock \emph{\bibinfo{title}{Geometry of Quantum States: An Introduction to
  Quantum Entanglement}} (\bibinfo{publisher}{Cambridge University Press},
  \bibinfo{address}{Cambridge, UK}, \bibinfo{year}{2006}).

\bibitem{AliCHT09}
\bibinfo{author}{Ali, S.~T.}, \bibinfo{author}{Carmeli, C.},
  \bibinfo{author}{Heinosaari, T.} \& \bibinfo{author}{Toigo, A.}
\newblock \bibinfo{title}{Commutative {POVM}s and fuzzy observables}.
\newblock \emph{\bibinfo{journal}{Found. Phys.}} \textbf{\bibinfo{volume}{39}},
  \bibinfo{pages}{593--612} (\bibinfo{year}{2009}).

\bibitem{Zami98}
\bibinfo{author}{Zamir, R.}
\newblock \bibinfo{title}{A proof of the {Fisher} information inequality via a
  data processing argument}.
\newblock \emph{\bibinfo{journal}{IEEE Trans. Inf. Theory}}
  \textbf{\bibinfo{volume}{44}}, \bibinfo{pages}{1246--1250}
  (\bibinfo{year}{1998}).

\bibitem{Zhu14IOC}
\bibinfo{author}{Zhu, H.}
\newblock \bibinfo{title}{Quantum state estimation with informationally
  overcomplete measurements}.
\newblock \emph{\bibinfo{journal}{Phys. Rev. A}} \textbf{\bibinfo{volume}{90}},
  \bibinfo{pages}{012115} (\bibinfo{year}{2014}).

\bibitem{DurtEBZ10}
\bibinfo{author}{Durt, T.}, \bibinfo{author}{Englert, B.-G.},
  \bibinfo{author}{Bengtsson, I.} \& \bibinfo{author}{{\.{Z}}yczkowski, K.}
\newblock \bibinfo{title}{On mutually unbiased bases}.
\newblock \emph{\bibinfo{journal}{Int. J. Quant. Inf.}}
  \textbf{\bibinfo{volume}{8}}, \bibinfo{pages}{535} (\bibinfo{year}{2010}).

\bibitem{Hall95}
\bibinfo{author}{Hall, M. J.~W.}
\newblock \bibinfo{title}{Information exclusion principle for complementary
  observables}.
\newblock \emph{\bibinfo{journal}{Phys. Rev. Lett.}}
  \textbf{\bibinfo{volume}{74}}, \bibinfo{pages}{3307--3311}
  (\bibinfo{year}{1995}).

\bibitem{ColeKW14}
\bibinfo{author}{Coles, P.~J.}, \bibinfo{author}{Kaniewski, J.} \&
  \bibinfo{author}{Wehner, S.}
\newblock \bibinfo{title}{Equivalence of wave--particle duality to entropic
  uncertainty}.
\newblock \emph{\bibinfo{journal}{Nat. Commun.}} \textbf{\bibinfo{volume}{5}},
  \bibinfo{pages}{5814} (\bibinfo{year}{2014}).

\bibitem{Land87}
\bibinfo{author}{Landau, L.~J.}
\newblock \bibinfo{title}{On the violation of {Bell's} inequality in quantum
  theory}.
\newblock \emph{\bibinfo{journal}{Phys. Lett. A}}
  \textbf{\bibinfo{volume}{120}}, \bibinfo{pages}{54 -- 56}
  (\bibinfo{year}{1987}).

\end{thebibliography}

\begin{thebibliography}{10}
\expandafter\ifx\csname url\endcsname\relax
  \def\url#1{\texttt{#1}}\fi
\expandafter\ifx\csname urlprefix\endcsname\relax\def\urlprefix{URL }\fi
\providecommand{\bibinfo}[2]{#2}
\providecommand{\eprint}[2][]{\url{#2}}

\bibitem{Fish25}
\bibinfo{author}{Fisher, R.~A.}
\newblock \bibinfo{title}{Theory of statistical estimation}.
\newblock \emph{\bibinfo{journal}{Math. Proc. Cambr. Philos. Soc.}}
  \textbf{\bibinfo{volume}{22}}, \bibinfo{pages}{700--725}
  (\bibinfo{year}{1925}).

\bibitem{Cram46}
\bibinfo{author}{Cram\'er, H.}
\newblock \emph{\bibinfo{title}{Mathematical Methods of Statistics}}
  (\bibinfo{publisher}{Princeton University Press},
  \bibinfo{address}{Princeton, NJ}, \bibinfo{year}{1946}).

\bibitem{Rao45}
\bibinfo{author}{Rao, C.~R.}
\newblock \bibinfo{title}{Information and the accuracy attainable in the
  estimation of statistical parameters}.
\newblock \emph{\bibinfo{journal}{Bull. Calcutta Math. Soc.}}
  \textbf{\bibinfo{volume}{37}}, \bibinfo{pages}{81--91}
  (\bibinfo{year}{1945}).

\bibitem{Frie99book}
\bibinfo{author}{Frieden, B.~R.}
\newblock \emph{\bibinfo{title}{Physics from {Fisher} Information: A
  Unification}} (\bibinfo{publisher}{Cambridge University Press},
  \bibinfo{address}{Cambridge, UK}, \bibinfo{year}{1999}).

\bibitem{Frie04book}
\bibinfo{author}{Frieden, B.~R.}
\newblock \emph{\bibinfo{title}{Science from {Fisher} Information: A
  Unification}} (\bibinfo{publisher}{Cambridge University Press},
  \bibinfo{address}{Cambridge, UK}, \bibinfo{year}{2004}).

\bibitem{Fish22}
\bibinfo{author}{Fisher, R.~A.}
\newblock \bibinfo{title}{On the mathematical foundations of theoretical
  statistics}.
\newblock \emph{\bibinfo{journal}{Philos. Trans. R. Soc. Lond. A}}
  \textbf{\bibinfo{volume}{222}}, \bibinfo{pages}{309--368}
  (\bibinfo{year}{1922}).

\bibitem{Hels76book}
\bibinfo{author}{Helstrom, C.~W.}
\newblock \emph{\bibinfo{title}{Quantum Detection and Estimation Theory}}
  (\bibinfo{publisher}{Academic Press}, \bibinfo{address}{New York},
  \bibinfo{year}{1976}).

\bibitem{Hole82book}
\bibinfo{author}{Holevo, A.~S.}
\newblock \emph{\bibinfo{title}{Probabilistic and Statistical Aspects of
  Quantum Theory}} (\bibinfo{publisher}{North-Holland},
  \bibinfo{address}{Amsterdam}, \bibinfo{year}{1982}).

\bibitem{GillM00}
\bibinfo{author}{Gill, R.~D.} \& \bibinfo{author}{Massar, S.}
\newblock \bibinfo{title}{State estimation for large ensembles}.
\newblock \emph{\bibinfo{journal}{Phys. Rev. A}} \textbf{\bibinfo{volume}{61}},
  \bibinfo{pages}{042312} (\bibinfo{year}{2000}).

\bibitem{Zhu12the}
\bibinfo{author}{Zhu, H.}
\newblock \emph{\bibinfo{title}{Quantum State Estimation and Symmetric
  Informationally Complete {POM}s}}.
\newblock Ph.D. thesis, \bibinfo{school}{National University of Singapore}
  (\bibinfo{year}{2012}).
\newblock \bibinfo{note}{Available at
  \url{http://scholarbank.nus.edu.sg/bitstream/handle/10635/35247/ZhuHJthesis.pdf}.
  Date of access: 31/10/2012}.

\bibitem{Hels67}
\bibinfo{author}{Helstrom, C.~W.}
\newblock \bibinfo{title}{Minimum mean-squared error of estimates in quantum
  statistics}.
\newblock \emph{\bibinfo{journal}{Phys. Lett. A}}
  \textbf{\bibinfo{volume}{25}}, \bibinfo{pages}{101--102}
  (\bibinfo{year}{1967}).

\bibitem{BrauC94}
\bibinfo{author}{Braunstein, S.~L.} \& \bibinfo{author}{Caves, C.~M.}
\newblock \bibinfo{title}{Statistical distance and the geometry of quantum
  states}.
\newblock \emph{\bibinfo{journal}{Phys. Rev. Lett.}}
  \textbf{\bibinfo{volume}{72}}, \bibinfo{pages}{3439--3443}
  (\bibinfo{year}{1994}).

\bibitem{Petz96}
\bibinfo{author}{Petz, D.}
\newblock \bibinfo{title}{Monotone metrics on matrix spaces}.
\newblock \emph{\bibinfo{journal}{Linear Algebra Appl.}}
  \textbf{\bibinfo{volume}{244}}, \bibinfo{pages}{81--96}
  (\bibinfo{year}{1996}).

\bibitem{PetzS96}
\bibinfo{author}{Petz, D.} \& \bibinfo{author}{Sud\'ar, C.}
\newblock \bibinfo{title}{Geometries of quantum states}.
\newblock \emph{\bibinfo{journal}{J. Math. Phys.}}
  \textbf{\bibinfo{volume}{37}}, \bibinfo{pages}{2662--2673}
  (\bibinfo{year}{1996}).

\bibitem{BengZ06book}
\bibinfo{author}{Bengtsson, I.} \& \bibinfo{author}{{\.{Z}}yczkowski, K.}
\newblock \emph{\bibinfo{title}{Geometry of Quantum States: An Introduction to
  Quantum Entanglement}} (\bibinfo{publisher}{Cambridge University Press},
  \bibinfo{address}{Cambridge, UK}, \bibinfo{year}{2006}).

\bibitem{BrauCM96}
\bibinfo{author}{Braunstein, S.~L.}, \bibinfo{author}{Caves, C.~M.} \&
  \bibinfo{author}{Milburn, G.~J.}
\newblock \bibinfo{title}{Generalized uncertainty relations: Theory, examples,
  and {Lorentz} invariance}.
\newblock \emph{\bibinfo{journal}{Ann. Phys.}} \textbf{\bibinfo{volume}{247}},
  \bibinfo{pages}{135 -- 173} (\bibinfo{year}{1996}).

\bibitem{Zhu14IOC}
\bibinfo{author}{Zhu, H.}
\newblock \bibinfo{title}{Quantum state estimation with informationally
  overcomplete measurements}.
\newblock \emph{\bibinfo{journal}{Phys. Rev. A}} \textbf{\bibinfo{volume}{90}},
  \bibinfo{pages}{012115} (\bibinfo{year}{2014}).

\bibitem{ZhuE11}
\bibinfo{author}{Zhu, H.} \& \bibinfo{author}{Englert, B.-G.}
\newblock \bibinfo{title}{Quantum state tomography with fully symmetric
  measurements and product measurements}.
\newblock \emph{\bibinfo{journal}{Phys. Rev. A}} \textbf{\bibinfo{volume}{84}},
  \bibinfo{pages}{022327} (\bibinfo{year}{2011}).

\end{thebibliography}

\section*{Acknowledgements}
It is a pleasure to thank Marcus Appleby, Jean-Daniel Bancal, Ingemar Bengtsson,  Giulio Chiribella,  Patrick  Coles, Lucien Hardy, Masahito Hayashi, Teiko Heinosaari, Ravi Kunjwal,  Tomasz Paterek, Valerio Scarani, Robert Spekkens,  Jun Suzuki, Daniel Terno, and Karol \.{Z}yczkowski for comments and discussions.  This research  was supported in part by Perimeter Institute for Theoretical Physics. Research at Perimeter Institute is supported by the Government of Canada through Industry Canada and by the Province of Ontario through the Ministry of Research and Innovation.

\end{cbunit}

\newpage

\begin{cbunit}
\begin{center}
\textbf{\large Supplementary information---Information complementarity: A new paradigm for decoding quantum incompatibility}
\end{center}

\setcounter{equation}{0}
\setcounter{figure}{0}
\setcounter{table}{0}
\makeatletter
\renewcommand{\theequation}{S\arabic{equation}}
\renewcommand{\thefigure}{S\arabic{figure}}

In this supplementary information we provide brief introduction on Fisher
information, Cram\'er--Rao bound, and quantum estimation theory. The concepts
of quantum Fisher information,  symmetric logarithmic derivative (SLD), quantum
Cram\'er--Rao bound, Gill--Massar inequality, and Gill Massar bound are reviewed.
 We also determine the complementarity chamber of the qubit based on the
 Gill--Massar inequality. In addition, we introduce parameter-free formulations
of the SLD bound and the Gill--Massar inequality, which are useful to studying
incompatibility criteria and measures.

\section{\label{sec:Fisher}Fisher information}
The Fisher information \cite{Fish25}  quantifies
the amount of information provided by an observation or a measurement
concerning certain parameters of interest.  It determines the minimal error
achievable in estimating  these parameters through the Cram\'er--Rao bound
\cite{Cram46, Rao45}. It is a
basic tool in statistical inference and also plays crucial roles in various
branches of physics and science in general \cite{Frie99book,Frie04book}.
Here our interest in Fisher information stems from its potential applications
in understanding a number of foundational issues in quantum mechanics, as
presented in the main text.

Consider a family of probability distributions $p(\xi|\theta)$
parametrized  by  $\theta$. Our task is to estimate the value of
$\theta$ as accurately as possible based on the measurement
outcomes. Given an outcome $\xi$, the probability $p(\xi|\theta)$ considered
as a function of
$\theta$ is called the \emph{likelihood function}. The \emph{score} is
defined as the partial derivative of the log-likelihood function
with respect to $\theta$ and reflects the sensitivity of the
log-likelihood function with respect to the variation of $\theta$.
Its first moment is zero, and the  second moment is known as the
\emph{Fisher information}
\cite{Fish25},
\begin{align}\label{sym:FisherInf}
I(\theta)&= \sum_\xi
p(\xi|\theta)\Bigl(\frac{\partial \ln p(\xi|\theta)}{\partial
\theta}\Bigr)^2
=\sum_\xi\frac{1}{p(\xi|\theta)}\Bigl(\frac{\partial
p(\xi|\theta)}{\partial \theta}\Bigr)^2.
\end{align}
The Fisher information represents the average sensitivity of the
log-likelihood function with respect to the variation of $\theta$.
Intuitively, the larger the Fisher information, the better we can
estimate the value of the parameter $\theta$.

An estimator $\hat{\theta}(\xi)$ of the parameter $\theta$ is
\emph{unbiased} if its expectation value is equal to the true
parameter; that is,
\begin{equation}
\sum_\xi p(\xi|\theta) [\hat{\theta}(\xi)-\theta]=0.
\end{equation}
In that case the variance or mean square error (MSE) of the estimator is
lower bounded by the inverse of the Fisher information, which is known as
the \emph{Cram\'er--Rao
bound} \cite{Cram46, Rao45}.

In the multiparameter setting, the Fisher information takes on  a matrix
form,
\begin{equation}
\begin{aligned}
I_{jk}(\theta)&=\sum_\xi p(\xi|\theta)\frac{\partial \ln
p(\xi|\theta)}{\partial \theta_j}\frac{\partial \ln
p(\xi|\theta)}{\partial \theta_k}.
\end{aligned}
\end{equation}
Accordingly, the Cram\'er--Rao bound for any unbiased estimator turns out
to be a matrix
inequality. Thanks to Fisher's theorem \cite{Fish22,Fish25}, the
lower bound can be saturated asymptotically with the maximum likelihood
estimator under very general assumptions.

\section{\label{sec:QET}Quantum estimation theory}
Here  we give a short introduction to quantum estimation theory tailored
to the needs in the main text. More details can be found in  \rscite{Hels76book,
Hole82book,GillM00,Zhu12the}.

In quantum parameter estimation, we are interested in the parameter
that characterizes the state $\rho(\theta)$ of a quantum system.  To
estimate the value of this parameter, we may perform generalized
measurements. Given a measurement $\Pi$ with outcomes $\Pi_\xi$, the
probability of obtaining the outcome $\xi$ is
$p(\xi|\theta)=\tr\{\rho(\theta)\Pi_\xi\}$. The corresponding Fisher
information $I_\Pi(\theta)$ reads
\begin{equation}
I_\Pi(\theta)=\sum_\xi
\frac{1}{p(\xi|\theta)}\tr\biggl\{\frac{\rmd
\rho(\theta)}{\rmd \theta}\Pi_\xi\biggr\}^2.
\end{equation}
Once a measurement is chosen,  the inverse Fisher information
sets a lower bound for the MSE of any unbiased estimator, which can be saturated
asymptotically by the maximum likelihood  estimator, as in the
case of classical parameter estimation.
It should be noted that the
bound depends on the specific measurement.

\subsection{Quantum Fisher information}
A measurement independent bound for the MSE can be derived based on  the
\emph{quantum  Fisher information}~\cite{Hels67,Hels76book, Hole82book}:
\begin{equation}\label{sym:QFISLD}
J(\theta)=\tr\{\rho(\theta) L(\theta)^2\},
\end{equation}
where $L(\theta)$  satisfies  the equation
\begin{equation}
\frac{\rmd \rho(\theta)}{\rmd \theta}=\frac{1}{2}[\rho(\theta)
L(\theta)+L(\theta)\rho(\theta)]
\end{equation}
and is known as  the \emph{symmetric logarithmic derivative} (SLD) of
$\rho(\theta)$ with respect to $\theta$.
The quantum  Fisher information $J(\theta)$ is a  an  upper bound for
the Fisher information $I(\theta)$, which is referred to as the SLD bound
henceforth. The bound can be saturated by measuring the observable $L(\theta)$.
Therefore, in the one-parameter setting, the complementarity chamber $\mathscr{C}(\theta)$
is a line segment determined by the equation $0\leq I(\theta)\leq J(\theta)$.
In conjunction with the classical Cram\'er--Rao bound, the inverse quantum
Fisher information
sets a lower bound for the MSE of any unbiased estimator, which is known
as the quantum Cram\'er--Rao bound~\cite{Hels67,Hels76book, Hole82book}.
In this paper, we are more concerned with the SLD bound $I(\theta)\leq J(\theta)$
itself rather than the bound for the MSE.

In addition to its application in quantum estimation theory, the
quantum Fisher information also plays an important role in studying
the geometry of quantum states~\cite{BrauC94,Petz96,PetzS96,BengZ06book}.
For example,  the SLD quantum
Fisher information allows  defining a statistical metric in the
state space that  is equal to four times of  the Bures metric~\cite{BrauC94}
and  generalizes the Fisher--Rao metric defined on the probability simplex~\cite{Fish25,Rao45,BengZ06book}.
With respect to this metric, the Bloch ball is a 3-hemisphere. Also, the
SLD quantum Fisher information plays a crucial role in studying parameter-based
uncertainty relations~\cite{BrauCM96}.

In the multiparameter setting both the Fisher information and the quantum
Fisher information take on  matrix form,
\begin{equation}
\begin{aligned}
I_{\Pi,jk}(\theta)&=\sum_\xi
\frac{1}{p(\xi|\theta)}\tr\bigl\{\rho_{,j}\Pi_\xi\bigr\}
\tr\bigl\{\rho_{,k}\Pi_\xi\bigr\},\\
J_{jk}(\theta)&=\frac{1}{2}\tr\bigl\{\rho(L_jL_k+L_kL_j)\bigr\},
\end{aligned}
\end{equation}
where $\rho_{,j}=\partial
\rho(\theta)/\partial \theta_j$ and  $L_j$ is the SLD associated with the
parameter $\theta_j$. As in the one-parameter
setting, $J(\theta)$ is an upper bound for $I(\theta)$. However, the  bound
generally cannot be saturated except when the $L_j$ can be measured simultaneously.
Consequently, the complementarity chamber is usually a small subset of the
set of hypothetical Fisher information matrices satisfying the SLD bound.
This difference is the main reason why multiparameter quantum estimation
problems are  so difficult  and poorly understood. Surprisingly, however,
this distinction  can also be turned into a powerful tool for studying the
complementarity principle, uncertainty relations and, in particular, the
joint measurement problem, which are the focus of the main text.

\subsection{Gill--Massar inequality  }
To better characterize the complementarity chamber in the multiparameter
setting, we need more powerful tools than the
SLD bound. One important tool is the following inequality derived by Gill
and Massar~\cite{GillM00} in the context of quantum state estimation,
\begin{equation}
\tr\{J^{-1}(\theta) I(\theta)\}\leq d-1,
\end{equation}
which is applicable to any measurement  on a $d$-level system.
The upper bound is saturated for any rank-one  measurement
when the number of parameters to be estimated is equal to the dimension $d^2-1$
 of the state space. The Gill--Massar (GM) inequality  succinctly
summarizes the information trade-off among incompatible observables in multiparameter
quantum estimation problems.
It sets a lower bound for the
weighted mean square error (WMSE) of any unbiased estimator~\cite{GillM00,
Zhu12the},
\begin{equation}\label{eq:GMboundWMSE}
\mathcal{E}_{W}^{\mathrm{GM}}=\frac{\bigl(\tr\sqrt{J^{-1/2}WJ^{-1/2}}\,\bigr)^2}{d-1},
\end{equation}
where $W$ is the weighting matrix (to simplify the notation we have omitted
the dependence on the parameter~$\theta$).
The lower bound can be saturated if and only if the hypothetical Fisher information
matrix
\begin{equation}\label{eq:FisherGM}
I_W=(d-1)J^{1/2}\frac{\sqrt{J^{-1/2}WJ^{-1/2}}}{\tr\sqrt{J^{-1/2}WJ^{-1/2}}}J^{1/2}
\end{equation}
belongs to the complementarity chamber. For example, the weighting matrix
for the mean square Bures distance is equal to one fourth of the quantum
Fisher information matrix, and the GM bound is $(d+1)^2(d-1)/4$.  The bound
can be saturated if and only if the complementarity chamber $\mathscr{C}$
contains $J/(d+1)$.

Both the Fisher information and quantum Fisher information depend on the
parametrization of the state space; a judicial choice is often crucial to
simplifying the discussion. For example, with a suitable parametrization,
we can turn the quantum Fisher information matrix into the identity at least
for a particular parameter point, say, $\tilde{\theta}$. Then the SLD bound
and the GM inequality reduce to $I(\tilde{\theta})\leq 1$ and $\tr\{I(\tilde{\theta})\}\leq
d-1$, respectively.

\subsection{Complementarity chamber for the qubit}
In the case of a qubit, the GM bound  for the  WMSE  can always be saturated,
and the GM inequality is both necessary and sufficient for characterizing
the complementarity chamber. Moreover, any Fisher information matrix saturating
the GM inequality can be realized by three mutually unbiased measurements.
To verify this claim, note that the inverse quantum Fisher information matrix
reads
$J^{-1}(\vec{s}) =1-\vec{s}\vec{s}$ in terms of the components of the Bloch
vector $\vec{s}$.
Suppose that
$I_W$ in \eref{eq:FisherGM} has eigenvalues $a_1, a_2, a_3$ along with orthonormal
eigenvectors $\vec{r}_1, \vec{r}_2,
\vec{r}_3$.  Denote by $s_1, s_2, s_3$ the three components
of the Bloch vector  in this basis. Then the GM
bound can be saturated by measuring each observable $\sigma_j:=\vec{r}_j\cdot\vec{\sigma}$
with
probability $a_j(1-s_j^2)$. Note that  the probabilities are normalized since
$\sum_ja_j(1-s_j^2)=\tr(J^{-1} I_W)=1$.
Therefore, the desired measurement scheme  can always be realized with a
complete set of mutually unbiased
measurements as claimed.

Alternatively, the structure of the complementarity chamber can be understood
by analogy as in the main text. For simplicity, we shall focus on the parameter
point
 $s=0$; the general situation can be analyzed along the same line of thinking.
 Since   $J=1$ at $s=0$, the set of Fisher information matrices saturating
the GM inequality
is isomorphic to the state space of  the three-dimensional real Hilbert space.
The extremal points of this set correspond to pure states, which form a real
projective space of dimension two. Each extremal Fisher information matrix
can be realized by a von Neumann measurement. A generic Fisher information
matrix in this set can be expressed as a convex combination of three extremal
Fisher information matrices, in analogy with the spectral decomposition of
the corresponding state. Note that the von Neumann measurements realizing
the three extremal Fisher information matrices are mutually unbiased. This
observation confirms the same conclusion as in the previous paragraph. It
should be noted that different convex decompositions of the given Fisher
information matrix may lead to different realizations.

\section{\label{sec:SLDGMT}Parameter-free formulations of the SLD bound and
the Gill--Massar inequality}
The SLD bound and GM inequality can  be formulated in a way that is parameter
free~\cite{Zhu12the}.  Such formulations are often much easier to work with
than the usual formulation and are quite useful in studying quantum estimation
theory. They are  particularly convenient to the current study since we are
more interested in measurements rather than states.  To derive such formulations,
we need to recast the Fisher information matrix and quantum Fisher information
matrix into superoperators.

\subsection{\label{sec:SLD}SLD bound}
Following the convention in~\rscite{Zhu12the,Zhu14IOC},
the Hilbert--Schmidt inner product between two operators $A$ and $B$ is denoted
by  $\dinner{A}{B}:=\tr(A^\dag B)$,  where the double ket notation is used
to distinguish them from ordinary kets. Given the state $\rho$ and measurement
$\Pi$, let $p_\xi=\tr(\rho\Pi_\xi)$ and $\bar{\Pi}_\xi=\Pi_\xi-\tr(\Pi_\xi)/d$.
Let $\mathbf{I}$ denote the identity superoperator and $\bid$  the projector
onto the space of traceless Hermitian operators.
Define
\begin{equation}\label{eq:FrameSOopt}
\begin{aligned}
\mathcal{F}(\rho)&:=\sum_{\xi}\dket{\Pi_\xi}\frac{1}{p_\xi}\dbra{\Pi_\xi},\\
\barcal{F}(\rho)&:=\bid
\mathcal{F}(\rho)\bid=\sum_{\xi}\dket{\bar{\Pi}_\xi}\frac{1}{p_\xi}\dbra{\bar{\Pi}_\xi},
\end{aligned}
\end{equation}
where the dependence on $\Pi$ is suppressed to simplify the notation.  Then
the Fisher information matrix can be written as
\begin{equation}\label{eq:FisherIMalternvative}
I_{jk}(\theta)=\dbra{\rho_{,j}}\mathcal{F}(\rho)\dket{\rho_{,k}}=\dbra{\rho_{,j}}\barcal{F}(\rho)\dket{\rho_{,k}}.
\end{equation}
Therefore, $\barcal{F}(\rho)$ is essentially the Fisher information matrix
in disguise~\cite{Zhu12the,Zhu14IOC}.

Define   superoperator $\mathcal{R}(\rho)$~\cite{BrauC94,Petz96,PetzS96}
by the
equation
\begin{equation}\label{eq:LRmultiplication}
\mathcal{R}(\rho)\dket{A}=\frac{1}{2}\dket{A\rho+\rho A}.
\end{equation}
Alternatively, $\mathcal{R}(\rho)$ can be written as
\begin{equation}
\mathcal{R}(\rho)=\frac{1}{2}\sum_{j,k=1}^d
\bigl(\dket{E_{jl}}\rho_{jk}\dbra{E_{kl}}+\dket{E_{lk}}\rho_{jk}\dbra{E_{lj}}\bigr),
\end{equation}
where   the $E_{jk}:=\outer{j}{k}$ form an  operator basis.
Define
\begin{equation}\label{sym:QFISLDso}
\mathcal{J}(\rho)=\mathcal{R}^{-1}(\rho), \quad
\barcal{J}(\rho)=\bid\mathcal{J}(\rho)\bid.
\end{equation}
Then we have
\begin{equation}\label{eq:QFIalternvative}
J_{jk}(\theta)=\dbra{\rho_{,j}}\mathcal{J}(\rho)\dket{\rho_{,k}}
=\dbra{\rho_{,j}}\barcal{J}(\rho)\dket{\rho_{,k}}.
\end{equation}
Therefore, $\barcal{J}(\rho)$ is the superoperator analogy of the quantum
Fisher information matrix.

Combining  \esref{eq:FisherIMalternvative} and~\eqref{eq:QFIalternvative},
we recognize that
the SLD bound for the Fisher information can be recast as
\begin{equation}\label{eq:SLDdouble}
\barcal{F}(\rho)\leq
\barcal{J}(\rho).
\end{equation}

\subsection{Gill--Massar inequality}
To derive alternative formulations of the GM inequality, we first note that
the GM trace $\tr\{J^{-1}(\theta) I(\theta)\}$ is independent of the parametrization
as long as the space spanned by the $\rho_{,j}$ is invariant. Let $\mathcal{P}$
be the projector onto this space, then
\begin{align}
\tr\{J^{-1}(\theta) I(\theta)\}&=\Tr\{[\mathcal{P} \mathcal{J}(\rho)\mathcal{P}]^+
\mathcal{F(\rho)}\}
=\Tr\{[\mathcal{P} \barcal{J}(\rho)\mathcal{P}]^+ \barcal{F}(\rho)\},
\end{align}
where $A^+$ denotes the Moore-Penrose generalized inverse of $A$, which is
equal to the inverse on the support of $A$ when $A$ is Hermitian. In addition,
the GM trace is nondecreasing when  the number of parameters increases  or
the space spanned by the $\rho_{,j}$ expands. Therefore,
\begin{equation}
\tr\{J^{-1}(\theta) I(\theta)\}\leq \Tr\{\barcal{J}^+(\rho) \barcal{F}(\rho)\},
\end{equation}
where the inequality is saturated when the number of parameters is equal
to $d^2-1$ or, equivalently, $\mathcal{P}=\bid$.
Another crucial observation are the  equalities
\begin{equation}
\dbra{\rho}\mathcal{F}(\rho)\dket{\rho}=\sum_\xi\tr (\rho\Pi_\xi)=1
\end{equation}
and
 \begin{align}\label{eq:QFIinverse}
\barcal{J}^{+}(\rho)&=\mathcal{J}^{-1}(\rho)-\douter{\rho}{\rho}.
\end{align}
Consequently,
\begin{align}
\Tr\{\barcal{J}^{+}(\rho)\barcal{F}(\rho)\}&=\Tr\{\barcal{J}^{+}(\rho)\mathcal{F}(\rho)\}
=\Tr\{\mathcal{J}^{-1}(\rho)\mathcal{F}(\rho)\}-1,
\end{align}
Therefore,  the GM inequality  admits two equivalent formulations,
\begin{equation}\label{eq:GMinequalityDouble}
\Tr\{\barcal{J}^{+}(\rho)\barcal{F}(\rho)\}\leq d-1,\quad
\Tr\{\mathcal{J}^{-1}(\rho)\mathcal{F}(\rho)\}\leq d.
\end{equation}

The above formulations  also lead to a much simpler proof of the GM inequality
\cite{Zhu12the}, whose original proof is quite convoluted.
\begin{align}
\Tr\{\mathcal{J}^{-1}(\rho)\mathcal{F}(\rho)\}&=\sum_\xi\frac{\dbra{\Pi_\xi}\mathcal{J}^{-1}(\rho)\dket{\Pi_\xi}}{\dinner{\rho}{\Pi_\xi}}
=\sum_\xi\frac{\tr(\rho\Pi_\xi^2)}{\tr(\rho\Pi_\xi)} \leq\sum_\xi\tr(\Pi_\xi)=d.
\end{align}
The inequality is saturated if the measurement is rank one.

In the case $\rho=1/d$ and thus $\mathcal{R}(\rho)=\mathbf{I}/d$, the SLD
bound in \eref{eq:SLDdouble} reduces to
$\barcal{G}\leq \bid$,
 where $\barcal{G}$ is the metric-adjusted Fisher information matrix  in
superoperator form, also known as the frame superoperator~\cite{ZhuE11,Zhu12the,Zhu14IOC},
\begin{equation}\label{eq:FrameSO-G}
\barcal{G}:=\bid \mathcal{G} \bid, \quad \mathcal{G}:=\frac{1}{d}\mathcal{F}\Bigl(\frac{1}{d}\Bigr)=\sum_{\xi}\dket{\Pi_\xi}\frac{1}{\tr\Pi_\xi}\dbra{\Pi_\xi}.
\end{equation}
Accordingly, the GM inequalities in \eref{eq:GMinequalityDouble} reduce to
\begin{equation}
\Tr(\barcal{G})\leq d-1,\quad   \Tr(\mathcal{G})\leq d.
\end{equation}
In this special case, the GM inequalities  are manifestly unitarily invariant.

\end{cbunit}

\end{document}